# STEP: Efficient Carbon Capture and S̲olar T̲hermal E̲lectrochemical P̲roduction of ammonia, fuels, cement, carbon nanotubes, metals and bleach

Stuart Licht[z]

Department of Chemistry, George Washington University, Washington, DC 20051, USA

Abstract

STEP (Solar Thermal Electrochemical Production) is an alternative solar energy conversion process. New and original, unpublished STEP results are compared with other STEP results. The STEP process uses semiconductors, solar energy and electrochemistry to generate a wide range of useful chemicals, rather than electricity, as the product. Using both subgap (to generate heat) and super bandgap (to generate electrons) insolation, STEP is more efficient than either photovoltaic or photoelectochemical solar energy conversion. STEP theory is derived and experimentally verified for the electrosynthesis of energetic molecules at high solar energy efficiency. In STEP the efficient formation of metals, fuels, chlorine, and carbon capture is driven by solar thermal heated electrolyses occuring at voltage below that of the room temperature energy stored in the products. As one example, $CO_2$ is reduced to either fuels, or storable carbon, at solar efficiency over 50% due to a synergy of efficient solar thermal absorption and electrochemical conversion at high temperature and reactant concentration. $CO_2$ is efficiently transformed to carbon nanotubes (C2CNT) with or without solar energy. New results on $CO_2$-free STEP ammonia, iron and cement production are delineated. Water is efficiently split to $H_2$ by molten electrolysis. A pathway is provided for the STEP decrease of atmospheric $CO_2$ levels to pre-industial levels in 10 years.

[z]E-mail: slicht@gwu.edu

1. Introduction

*Anthropogenic release of carbon dioxide and atmospheric carbon dioxide have reached record levels.* One path towards $CO_2$ reduction is to utilize renewable energy to produce electricity. Another, less explored, path is to utilize renewable energy to directly produce societal staples such as metals, bleach, fuels, including carbonaceous fuels. Whereas solar driven water splitting to generate hydrogen fuels has been extensively studied,[1] there are fewer studies of solar driven carbon dioxide splitting. Here, new, unpublished original STEP results are compared with our earlier other STEP results.[2] We introduced a global process for the Solar Thermal Electrochemical Production (STEP) of energetic molecules, including $CO_2$ splitting.[3-8] *$CO_2$ is a highly stable, noncombustible molecule, and its thermodynamic stability makes its activation energy demanding and challenging.*"[9] "In search of a solution for climate change associated with increasing levels of atmospheric $CO_2$, the field of carbon dioxide splitting (solar or otherwise), while young, is growing rapidly, and as with water splitting, includes the study of photoelectrochemical, biomimetic, electrolytic, and thermal pathways of carbon dioxide splitting.[10,11]

The direct thermal splitting of $CO_2$ requires excessive temperatures to drive any significant dissociation. As a result, lower temperature thermochemical processes using coupled reactions have recently been studied.[12-16] The coupling of multiple reactions steps decreases the system efficiency. To date, such challenges, and the associated efficiency losses, have been an impediment to the implementation of the related, extensively studied field of thermochemical splitting of water.[2] Photoelectrochemistry probes the energetics of illuminated semiconductors in an electrolyte, and provides an alternative path to solar fuel formation. Photoelectrochemical solar cells (**PEC**s) can convert solar energy to electricity,[17-21] and with inclusion of an electrochemical storage couple, have the capability for internal energy storage, to provide a level output despite variations in sunlight.[22,23] Solar to photoelectrochemical energy can also be stored externally in chemical form, when it is used to drive the formation of energetically rich chemicals. Photochemical, and photoelectrochemical, splitting of carbon dioxide [24-29] have demonstrated selective production of specific fuel products. Such systems function at low current density and efficiencies of ~1 percent, and as with photoelectrochemical water splitting face stability and bandgap challenges related to effective operation with visible light.[21,30,31]

The electrically driven (nonsolar) electrolysis of dissolved carbon dioxide is under investigation at or near room temperature in aqueous, non-aqueous and PEM media.[32-41] These are constrained by the thermodynamic and kinetic challenges associated with ambient temperature, endergonic processes, of a high electrolysis potential, large overpotential, low rate and low electrolysis efficiency. High temperature, solid oxide electrolysis of carbon dioxide dates back to 1960 suggestions to use such cells to renew air for a space habitat,[42-44] and the sustainable rate of the solid oxide reduction of carbon dioxide is improving rapidly.[45-51] Molten carbonate, rather solid oxide, fuel cells running in the reverse mode had also been studied to renew air in 2002.[52] In a manner analogous to our 2002 high temperature solar water splitting studies (described below),[53-56] we showed in 2009 that molten carbonate cells are particularly effective for the solar driven electrolysis of carbon dioxide,[3,4,8] and also $CO_2$-free iron metal production.[5,6]

Light driven water splitting was originally demonstrated with $TiO_2$ (a seminconductor with a bandgap, $E_g > 3.0$ eV).[57] However, only a small fraction of sunlight has sufficient energy to drive $TiO_2$ photoexcitation. Studies had sought to tune (lower) the semiconductor bandgap to provide a better match to the electrolysis potential.[58] In 2000, we used external multiple bandgap **PV**s (photovoltaics) to generate $H_2$ by splitting water at 18% solar energy conversion efficiency.[59,60] However, that room temperature process does not take advantage of additional, available thermal energy.

An alternative to tuning a seminconductor bandgap to provide a better match to the solar spectrum, is an approach to tune (lower) the electrolysis potential.[53-55] In 2002 we introduced a photo electrochemical thermal water splitting theory,[53,54] which was verified by experiment in 2003, for $H_2$ generation at over 30% solar energy conversion efficiency, and providing the first experimental demonstration that a semiconductor, such as Si ($E_g$ = 1.1eV), with bandgap lower than the standard water splitting potential ($E°_{H_2O}$(25°C) =1.23 V), can directly drive hydrogen formation.[55] With increasing temperature, the quantitative decrease in the electrochemical potential to split water to hydrogen and oxygen had been well known by the 1950's.[61a,61b] In 1976 Wentworth and Chen wrote on "simple thermal decomposition reactions for storage of solar energy," with the limitation that the products of the reaction must be separated to prevent back reaction (and without any electrochemical component),[62] and as early as 1980 it was noted that thermal energy could decrease the necessary energy for the generation of $H_2$ by electrolysis.[63] However, the process combines elements of solid state physics, insolation and electrochemical theory, complicating rigorous theoretical support of the process. Our photo electrochemical thermal water splitting model for solar/$H_2$ by this process, was the first derivation of bandgap restricted, thermal enhanced, high solar water splitting efficiencies. The model, predicting solar energy conversion efficiencies that exceed those of conventional photovoltaics was initially derived for AM(Air Mass)1.5, terrestrial insolation, and later expanded to include sunlight above the atmosphere (AM0 insolation).[53,54] The experimental accomplishment followed, and established that the water splitting potential can be specifically tuned to match efficient photo-absorbers,[55,56] eliminating the challenge of tuning (varying) the semiconductor bandgap, and which can lead to over 30% solar to chemical energy conversion efficiencies. Our early process was specific to $H_2$ and did not incorporate the additional temperature enhancement of excess super-band gap energy and concentration enhancement of excess reactant to further decrease the electrolysis potential, in our contemporary **STEP** process.

## 2. Solar Thermal Electrochemical Production of Energetic Molecules: An Overview

### 2.1. STEP Theoretical Background

A single, small band gap junction, such as in a silicon PV, cannot generate the minimum photopotential required to drive many room temperature electrolysis reactions, as shown in the left of Scheme 1. The advancement of such studies had focused on tuning semiconductor bandgaps[58] to provide a better match to the electrochemical potential (specifically, the water splitting potential), or by utilizing more complex, multiple bandgap structures using multiple photon excitation.[59,60] Either of these structures are not capable of excitation beyond the bandedge and can not make use of longer wavelength sunlight. Photovoltaics are limited to super-bandgap sunlight, $h\nu > E_g$, precluding use of long wavelength radiation, $h\nu < E_g$. STEP instead directs this IR sunlight to heat electrochemical reactions, and uses visible sunlight to generate electronic charge to drive these electrolyses.

Rather than tuning the bandgap to provide a better energetic match to the electrolysis potential, the **STEP** process instead tunes the redox potential to match the bandgap. The right side of Scheme 1 presents the energy diagram of a STEP process. The high temperature pathway decreases the free energy requirements for processes whose electrolysis potential decreases with increasing temperature. STEP uses solar energy to drive, otherwise energetically forbidden, pathways of charge transfer. The process combines elements of solid state physics, insolation (solar illumination) and high temperature electrochemical energy conversion. Kinetics improve, and endergonic thermodynamic potentials, decrease with increasing temperature. The result is a synergy, making use of the full spectrum of sunlight, and capturing more solar energy. **STEP** is intrinsically more efficient than other solar energy conversion processes, as it utilizes not only the visible sunlight used to drive PVs, but also utilizes the previously detrimental (due to PV thermal degradation) thermal component of sunlight, for the electrolytic formation of chemicals.

The two bases for improved efficiencies using the STEP process are (i) excess heat, such as unused heat in solar cells, can be used to increase the temperature of an electrolysis cell, such as for electrolytic $CO_2$ splitting, while (ii) the product to reactant ratio can be increased to favor the kintetic and energetic formation of reactants. With increasing temperature, the quantitative decrease in the electrochemical potential to drive a variety of electrochemical syntheses is well known, substantially decreasing the electronic energy (the electrolysis potential) required to form energetic products. The extent of the decrease in the electrolysis potential, $E_{redox}$, may be tuned by choosing the constituents and temperature of the electrolysis. The process distinguishes radiation that is intrinsically energy sufficient to drive **PV** charge transfer, and applies all excess solar thermal energy to heat the electrolysis reaction chamber.

Scheme 2 summarizes the charge, heat and molecular flow for the **STEP** process; the high temperature pathway decreases the potential required to drive endergonic electrolyses, and also facilitates the kinetics of charge transfer (i.e., decreases overpotential losses), which arise during electrolysis. This process consists of (i) sunlight harvesting and concentration, (ii) photovoltaic charge transfer driven by super-bandgap energy, (iii) transfer of sub-bandgap and excess super-bandgap radiation to heat the electrolysis chamber, (iv) high temperature, low energy electrolysis forming energy rich products, and (v) cycle completion by pre-heating of the electrolysis reactant through heat exchange with the energetic electrolysis products. As indicated on the right side of Scheme 2, the light harvesting can use various optical configurations; e.g. in lieu of parabolic, or Fresnel, concentrators, a heliostat/solar tower with secondary optics can achieve higher process temperatures (>1000 °C) with concentrations of ~2000 suns. Beam splitters can redirect sub-bandgap radiation away from the PV (minimzing PV heating) for a direct heat exchange with the electrolyzer.

Solar heating can decrease the energy to drive a range of electrolyses. Such processes can be determined using available entropy, S, and enthalpy, H, and free-energy, G, data,[61b] and are identified by their negative isothermal temperature coefficient of the cell potential.[61a] This coefficient $(dE/dT)_{isoth}$ is the derivative of the electromotive force of the isothermal cell:

$$(dE/dT)_{isoth} = \Delta S/nF = (\Delta H - \Delta G)/nFT \qquad [1]$$

The starting process of modeling any STEP process is the conventional expression of a generalized electrochemical process, in a cell which drives an n electron charge transfer electrolysis reaction, comprising "x" reactants, $R_i$, with stoichiometric coefficients $r_i$, and yielding "y" products, $C_i$, with stoichiometric coefficients $c_i$.

Electrode 1 | Electrolyte | Electrode 2

Using the convention of $E = E_{cathode} - E_{anode}$ to describe the positive potential necessary to drive a non-spontaneous process, by transfer of n electrons in the electrolysis by transfer of n electrons in the electrolysis reaction of reactants to products:

$$\sum_{i=1 \text{ to } x} r_i R_i \rightarrow \sum_{i=1 \text{ to } y} c_i C_i \qquad [2]$$

At any electrolysis temperature, $T_{STEP}$, and at unit activity, the reaction has electrochemical potential, $E°_T$. This may be calculated from consistent, compiled unit activity thermochemical data sets, such as the NIST condensed phase and fluid properties data sets,[61b] as:

$$E°_T = -\Delta G°(T=T_{STEP})/nF; \quad E°_{ambient} \equiv E°_T(T_{ambient}); \text{ here } T_{ambient} = 298.15K = 25°C,$$

and: $\Delta G°(T=T_{STEP}) = \sum_{i=1 \text{ to } y} c_i(H°(C_i,T) - TS°(C_i,T)) - \sum_{i=1 \text{ to } x} r_i(H°(R_i,T) - TS°(R_i,T)) \qquad [3]$

Compiled thermochemical data are often based on different reference states, while a consistent reference state is needed to understand electrolysis limiting processes, including water.[64,65] This challenge is overcome by modification of the unit activity (a=1) consistent calculated electrolysis potential to determine the potential at other reagent and product relative activities via the Nernst equation.[66,67] Electrolysis provides control of the relative amounts of reactant and generated product in a system. A substantial activity differential can also drive **STEP** improvement at elevated temperature, and will be derived. The potential variation with activity, a, of the Equation 2 reaction is given by:

$$E_{T,a} = E°_T - (RT/nF) \cdot \ln( \prod_{i=1 \text{ to } x} a(R_i)^{r_i} / \prod_{i=1 \text{ to } y} a(C_i)^{c_i} ) \qquad [4]$$

Electrolysis systems with a negative isothermal temperature coefficient tend to cool as the electrolysis products are generated. Specifically in endergonic electrolytic processes, the eq 4 free-energy electrolysis potential, $E_T$, is less than the enthalpy based potential. This latter value is the potential at which the system temperature would remain constant during electrolysis. This thermoneutral potential, $E_{tn}$, is given by:

$$E_{tn}(T_{STEP}) = -\Delta H(T)/nF; \quad \Delta H(T_{STEP}) = \sum_{i=1 \text{ to } b} c_i H(C_i, T_{STEP}) - \sum_{i=1 \text{ to } a} r_i H(R_i, T_{STEP}) \qquad [5]$$

Two general STEP implementations are being explored. Both can provide the thermoneutral energy to sustain a variety of electrolyses. The thermoneutral potential, determined from the enthalpy of a reaction, describes the energy required to sustain an electrochemical process without cooling. For example, the thermoneutral potential we have calculated and reported for $CO_2$ splitting to CO and $O_2$ at unit activities, from eq 5, is 1.46($\pm$0.01) V over the temperature range of 25-1400°C. As represented in Scheme 3 on the left, the standard electrolysis potential at room temperature, E°, can comprise a significant fraction of the thermoneutral potential. The first STEP mode, energetically represented next to the room temperature process in the scheme, separates sunlight into thermal and visible radiation. The solar visible generates electronic charge which drives electrolysis charge transfer. The solar thermal component heats the electrolysis and decreases both the E° at this higher T, and the overpotential. The second mode, termed Hy-STEP (on the right) from "hybrid-STEP", does not separate sunlight, and instead directs all sunlight to heating the electrolysis, generating the highest T and smallest E, while the electrical energy for electrolysis is generated by a separate source (such as by photovoltaic, solar thermal electric, wind turbine, hydro, nuclear or fossil fuel generated electronic charge). As shown on the right side, high relative concentrations of the electrolysis reactant (such as $CO_2$ or iron oxide will further decrease the electrolysis potential).

2.2 STEP Solar to Chemical Energy Conversion Efficiency

The Hy-STEP mode is being studied outdoors with either wind or solar CPV (concentrator photovoltaic) generated electricity to drive $E_{electrolysis}$. The STEP mode is experimentally more complex and is presently studied indoors under solar simulator illumination. Determination of the efficiency of Hy-STEP with solar electric is straightforward in the domain in which $E_{electrolysis} < E_{thermoneutral}$ and the coulombic efficiency is high. Solar thermal energy is collected at an efficiency of $\eta_{thermal}$ to decrease the energy from $E_{thermoneutral}$ to $E_{electrolysis}$, and then electrolysis is driven at a solar electric energy efficiency of $\eta_{solar\text{-}electric}$:

$$\eta_{\text{Hy-STEP solar}} = (\eta_{thermal} \bullet (E_{thermoneutral}-E_{electrolysis}) + \eta_{solar\text{-}electric} \bullet E_{electrolysis}) / E_{thermoneutral} \qquad [6]$$

$\eta_{thermal}$ is higher than $\eta_{solar\text{-}electric}$, and gains in efficiency occur in eq 6 in the limit as $E_{electrolysis}$ approaches 0. $E_{electrolysis} = 0$ is equivalent to thermochemical, rather than electrolytic, production. As seen in Figure 1, at unit activity $E°_{CO2/CO}$ does not approach 0 until 3000°C. Material constraints inhibit approach to this higher temperature, while electrolysis also provides the advantage of spontaneous product seperation. At lower temperature, small values of $E_{electrolysis}$ can occur at higher reactant and lower product activities, as described in eq 4. In the present configuration sunlight is concentrated at 75% solar to thermal efficiency, heating the electrolysis to 950°C, which decreases the high current density $CO_2$ splitting potential to 0.9V, and the electrolysis charge is provided by CPV at 37% solar to electric efficiency. The solar to chemical energy conversion efficiency is in accordance with eq 6:

$$\eta_{\text{Hy-STEP solar}} = (75\% \bullet (1.46V-0.90V) + 37\% \bullet 0.90V)/1.46V = 52\% \qquad [7]$$

A relatively high concentration of reactants lowers the voltage of electrolysis via the Nernst term in eq 4. With appropriate choice of high temperature electrolyte, this effect can be dramatic, for example both in STEP iron and in comparing the benefits of the molten carbonate to solid oxide (gas phase) reactants for STEP $CO_2$ electrolytic reduction, sequestration and fuel formation. Fe(III) (as found in the common iron ore, hematite) is nearly insoluble in sodium carbonate, while it is soluble to over 10 m (molal) in lithium carbonate,[6] and as discussed in Section 2.3, molten carbonate electrolyzer provides $10^3$ to $10^6$ times higher concentration of reactant at the cathode surface than a solid oxide electrolyzer.

In practice, for STEP iron or carbon capture, we simultaneously drive lithium carbonate electrolysis cells together in series, at the CPV maximum power point (Figure 2). Specifically, a Spectrolab CDO-100-C1MJ concentrator solar cell is used to generate 2.7 V at maximum power point, with solar to electrical energy efficiencies of 37% under 500 suns illumination. As seen in Figure 2, at maximum power, the 0.99 $cm^2$ cell generates 1.3 A at 100 suns, and when masked to 0.2 $cm^2$ area generates 1.4 A at 500 suns. Electrolysis electrode surface areas were chosen to match the solar cell generated power. At 950°C at 0.9V, the electrolysis cells generate carbon monoxide at 1.3 to 1.5 A (the electrolysis current stability is shown at the bottom of Figure 2).

In accord with eq 6 and Scheme 3, Hy-STEP efficiency improves with temperature increase to decrease overpotential and $E_{electrolysis}$, and with increase in the relative reactant activity. Higher solar efficiencies will be expected, both with more effective carbonate electrocatalysts (as morphologies with higher effective surface area and lower overpotential) are developed, and as also as PV efficiencies increase. Increases in solar to electric (both PV, CPV and solar thermal-electric) efficiencies continue to be reported, and will improve eq 7 efficiency. For example, multijunction CPV have been reported improved to $\eta_{PV}$ = 40.7%.[71]

Engineering refinements will improve some aspects, and decrease other aspects, of the system efficiency. Preheating the $CO_2$, by circulating it as a coolant under the CPV (as we currently do in the indoor STEP experiment, but not outdoor, Hy-STEP experiments) will improve the system efficiency. In the present configuration outgoing CO and $O_2$ gases at the cathode and anode heat the incoming $CO_2$. Isolation of the electrolysis products will require heat exchangers with accompanying radiative heat losses, and for electrolyses in which there are side reactions or product recombination losses, $\eta_{\text{Hy-STEP solar}}$ will decrease proportional to the decrease in coulombic efficiency. At present, wind turbine generated electricity is more cost effective than solar-electric, and we have demonstrated a Hy-STEP process with wind-electric, for $CO_2$ free production of iron (delineated in Section 3.8). Addition of long-term (overnight) molten salt insulated storage will permit continuous operation of the STEP process. Both STEP implementations provide a basis for practical, high solar efficiencies.

Components for STEP $CO_2$ capture and conversion to solid carbon are represented on the left side of Figure 2, and are detailed in references 4-7. A 2.7 V CPV phootopotential drives three in series electrolyses at 950°C. Fundamental details of the heat balance are provided in reference 4. The CPV has an experimental solar efficiency of 37%, and the 63% of insolation not converted to electricity comprises a significant heat source. The challenge is to direct a substantial fraction of this heat to the electrolysis. An example of this challenge is in the first stage of heating, in which higher temperatures increases $CO_2$ preheat, but diminishes the CPV power. Heating of the reactant $CO_2$ is a three tier process in the current configuration: the preheating of room temperature $CO_2$ consists of either (1a) flow-through a heat exchange fixed to the back of the concentrator solar cell and/or (1b) preheating to simulate $CO_2$ extracted from an available heat source such as a hot smoke (flue) stack, (2) secondary heating consists of passing this $CO_2$ through a heat exchange with the outgoing products, (3) tertiary heat is applied through concentrated, split solar thermal energy (Figure 2).

An upper limit to the energy required to maintain a constant system temperature is given in the case in which neither solar IR, excess solar visible, nor heat exchange from the environment or products would be applied to the system. When an 0.90V electrolysis occurs, an additional 0.56 V, over $E_{tn}$ =1.46V , is required to maintain a constant system temperature. Hence, in the case of three electrolyses in series, as in Figure 2, an additional 3x0.56V=1.68V will maintain constant temperature. This is less than the 63% of the solar energy (equivalent to 4.6V) not used in generating the 2.7 V of maximum power point voltage of electronic charge from the CPV in this experiment. Heating requirements are even less, when the reactant activity is maintained at a level that is higher than the product activity. For example, this is accomplished when products are continuously removed to ensure that the partial pressure of the products is lower than that of the $CO_2$. This lowers the total heat required for temperature neutrality to below that of the unit activity thermoneutral potential 1.46V.

The STEP effective solar energy conversion efficiency, $\eta_{STEP}$, is constrained by both photovoltaic and thermal boost conversion efficiencies, $\eta_{PV}$ and $\eta_{thermal\text{-}boost}$.[8] Here, the CPV sustains a conversion efficiency of $\eta_{PV}$ = 37.0%. In the system, passage of electrolysis current requires an additional, combined (ohmic, & anodic + cathodic over-) potential above the thermodynamic potential. However, mobility and kinetics improve at higher temperature to decrease this overpotential. The generated CO contains an increase in oxidation potential compared to carbon dioxide at room temperature ($E_{CO2/CO}$(25°C)= 1.33 V for $CO_2 \rightarrow$ CO +1/2$O_2$ in Figure 1), an increase of 0.43 V compared to the 0.90 V used to generate the CO. The electrolysis efficiency compares the stored potential to the applied potential, $\eta_{thermal\text{-}boost}$ = $E°_{electrolysis}$(25°C) / $V_{electrolysis}$(T).[4] Given a stable temperature electrolysis environment, the experimental STEP solar to CO carbon capture and conversion efficiency is the product of this relative gain in energy and the electronic solar efficiency:

$$\eta_{STEP} = \eta_{PV} \cdot \eta_{thermal\text{-}boost} = 37.0\% \cdot (1.33V/0.90V) = 54.7\% \qquad [8]$$

Ohmic and overpotential losses are already included in the measured electrolysis potential. This 54.7% STEP solar conversion efficiency is an upper limit of the present experiment, and as with the Hy-STEP mode, improvements are expected in electrocatalysis and CPV efficiency. Additional losses will occur when beam splitter and secondary concentrator optics losses, and thermal systems matching are incorporated, but serves to demonstrate the synergy of this solar/photo/electrochemical/thermal process, leads to energy efficiency higher than that for solar generated electricity,[71] or for photochemical,[72] photoelectrochemical,[21,27] solar thermal,[73] or other $CO_2$ reduction processes.[74]

The CPV does not need, nor function with, sunlight of energy less than that of the 0.67 eV bandgap of the multi-junction Ge bottom layer. From our previous calculations, this thermal energy comprises 10% of AM1.5 insolation, which will be further diminished by the solar thermal absorption efficiency and heat exchange to the electrolysis efficiency,[54] and under 0.5 MW m$^{-2}$ of incident sunlight (500 suns illumination), yields ~50 kW m$^{-2}$, which may be split off as thermal energy towards heating the electrolysis cell without decreasing the CPV electronic power. The CPV, while efficient, utilizes less than half of the super-bandgap (hν > 0.67 eV) sunlight. A portion of this > ~250 kW m$^{-2}$ available energy, is extracted through heat exchange at the backside of the CPV. Another useful source for consideration as supplemental heat is industrial exhaust. The temperature of industrial flue stacks varies widely, with fossil fuel source and application, and ranges up to 650°C for an open circuit gas turbine. The efficiency of thermal energy transfer will limit use of this available heat.

A lower limit to the STEP efficiency is determined when no heat is recovered, either from the CPV or remaining solar IR, and when heat is not recovered via heat exchange from the electrolysis products, and when an external heat source is used to maintain a constant electrolysis temperature. In this case, the difference between the electrolysis potential and the thermoneutral potential represents the enthalpy required to keep the system from cooling. In this case, our 0.9V electrolysis occurs at an efficiency of (0.90V/1.46V) · 54.7% = 34%. While the STEP energy analysis, detailed in Section 4.2 for example for $CO_2$ to CO splitting, is more complex than that of the Hy-STEP mode, more solar thermal energy is available including a PV's unused or waste heat to drive the process and to improve the solar to chemical energy conversion efficiency. We determine the STEP solar efficiency over the range from inclusion of no solar thermal heat (based on the enthalpy, rather than free energy, of reaction) to the case where the solar thermal heat is sufficient to sustain the reaction (based on the free energy of reaction). This determines the efficiency range, as chemical flow out to the solar flow in (as measured by the increase in chemical energy of the products compared to the reactants), from 34% to over 50%.

2.3. Identification of STEP consistent endergonic processes

The electrochemical driving force for a variety of chemicals of widespread use by society will be shown to significantly decrease with increasing temperature. As calculated and summarized in the top left of Figure 1, the electrochemical driving force for electrolysis of either carbon dioxide or water, significantly decreases with increasing temperature. The ability to remove $CO_2$ from exhaust stacks or atmospheric sources, provides a response to linked environmental impacts, including global warming due to anthropogenic $CO_2$ emission. From the known thermochemical data for $CO_2$, CO and $O_2$, and in accord with eq 1, $CO_2$ splitting can be described by:

$CO_2(g) \rightarrow CO(g) + 1/2 O_2(g)$;

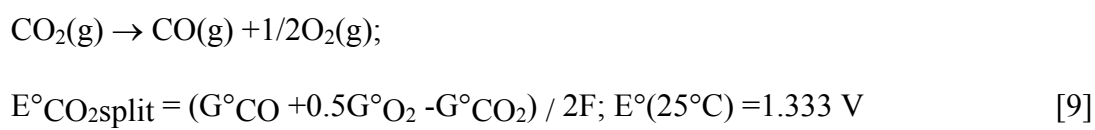

$E°_{CO_2split} = (G°_{CO} + 0.5 G°_{O_2} - G°_{CO_2}) / 2F; \; E°(25°C) = 1.333 \text{ V}$ [9]

As an example of the solar energy efficiency gains, this progress report focuses on $CO_2$ splitting potentials, and provides examples of other useful STEP processes. As seen in **Figure 1**, $CO_2$ splitting potentials decrease more rapidly with temperature than those for water splitting, signifying that the STEP process may be readily applied to $CO_2$ electrolysis. Efficient, renewable, non-fossil fuel energy rich carbon sources are needed, and the product of eq 9, carbon monoxide is a significant industrial gas with a myriad of uses, including the bulk manufacturing of hydrocarbon fuels, acetic acid and aldehydes (and detergent precursors), and for use in industrial nickel purification.[68] To alleviate challenges of fossil-fuel resource depletion, CO is an important syngas component and methanol is formed through the reaction with $H_2$. The ability to remove $CO_2$ from exhaust stacks or atmospheric sources, also limits $CO_2$ emission. Based on our original analogous experimental photo-thermal electrochemical water electrolysis design,[55,56] the first $CO_2$ STEP process consists of solar driven and solar thermal assisted $CO_2$ electrolysis. In particular, in a molten carbonate bath electrolysis cell, fed by $CO_2$.

cathode:   $2CO_2(g) + 2e^- \rightarrow CO_3^{2-}(molten) + CO(g)$

anode:   $CO_3^{2-}(molten) \rightarrow CO_2(g) + 1/2 O_2(g) + 2e^-$

cell:   $CO_2(g) \rightarrow CO(g) + 1/2 O_2(g)$   [10]

Molten alkali carbonate electrolyte fuel cells typically operate at 650°C. Li, Na or K cation variation can affect charge mobility and operational temperatures. Sintered nickel often serves as the anode, porous lithium doped nickel oxide often as the cathode, while the electrolyte is suspended in a porous, insulating, chemically inert $LiAlO_2$ ceramic matrix.[69]

Solar thermal energy can be used to favor the formation of products for electrolyses characterized by a negative isothermal temperature coefficient, but will not improve the efficiency of heat neutral or exothermic reactions. An example of this restriction occurs for the electrolysis reaction currently used by industry to generate chlorine. During 2008, the generation of chlorine gas (principally for use as bleach and in the chlor-alkali industry) consumed approximately 1% of the world's electricity,[70] prepared in accord with the industrial electrolytic process:

$2NaCl + 2H_2O \rightarrow Cl_2 + H_2 + 2NaOH$; E°(25°C) = 2.502 V   [11]

In the lower left portion of Figure 1, the calculated electrolysis potential for this industrial chlor-alkali reaction exhibits little variation with temperature, and hence the conventional generation of chlorine by electrolysis would not benefit from the inclusion of solar heating. This potential is relatively invariant, despite a number of phase changes of the components (indicated on the figure and which include the melting of NaOH or NaCl). However, as seen in the figure, the calculated potential for the anhydrous electrolysis of chloride salts is endergonic , including the electrolyses to generate not only chlorine, but also metallic lithium, sodium and magnesium, and can be greatly improved through the **STEP** process:

$MCl_n \rightarrow n/2 Cl_2 + M$; $E°_{MCl_n split}(25°C)$ =3.98 V for M=Li or Na; 4.24 V for M=K; 3.07 V for M=Mg [12]

The calculated decrease for the anhydrous chloride electrolysis potentials are on the order of volts per 1000°C temperature change. For example, from 25°C up to the $MgCl_2$ boiling point of 1412°C, the $MgCl_2$ electrolysis potential decreases from 3.07 V to 1.86 V. This decrease provides a theoretical basis for significant, non $CO_2$ emitting, non-fossil fuel consuming processes for the generation of chlorine and magnesium, to be delineated in Section 3.9, and occurring at high solar efficiency analogous to the similar $CO_2$ **STEP** process.

In Section 3.2 the **STEP** process will be derived for the efficient solar removal / recycling of $CO_2$. In addition, thermodynamic calculation of metal and chloride electrolysis rest potentials identifies electrolytic processes which are consistent with endergonic processes for the formation of iron, chlorine, aluminum, lithium, sodium and magnesium, via $CO_2$–free pathways. As shown, the conversion and replacement of the conventional, aqueous, industrial alkali-chlor process, with an anhydrous electrosynthesis, results in a redox potential with a calculated decrease of 1.1 V from 25°C to 1000°C.

As seen in the top right of Figure 1, the calculated electrochemical reduction of metal oxides can exhibit a sharp, smooth decrease in redox potential over a wide range of phase changes. These endergonic  process provide an opportunity for the replacement of conventional industrial processes by the **STEP** formation of these metals. In 2008, industrial electrolytic processes consumed ~5% of the world's electricity, including for aluminum (3%), chlorine (1%), and lithium, magnesium and sodium production. This 5% of the global $19 \times 10^{12}$ kWh of electrical production, is equivalent to the emission of $6 \times 10^8$ metric tons of $CO_2$.[70] The iron and steel industry accounts for a quarter of industrial direct $CO_2$ emissions. Currently, iron is predominantly formed through the reduction of hematite with carbon, emitting $CO_2$:

$$Fe_2O_3 + 3C + 3/2 O_2 \rightarrow 2Fe + 3CO_2 \qquad [13]$$

A non-$CO_2$ emitting alternative is provided by the **STEP** driven electrolysis of $Fe_2O_3$:

$$Fe_2O_3 \rightarrow 2Fe + 3/2 O_2 \qquad E° = 1.28 \text{ V} \qquad [14]$$

As seen in the top right of Figure 1, the calculated iron generating electrolysis potentials drops 0.5 V (a 38% drop) from 25°C to 1000 °C, and as with the $CO_2$ analogue, will be expected to decrease more rapidly with high iron oxide activity conditions. Conventional industrial processes for these metals and chlorine, along with $CO_2$ emitted from power and transportation, are responsible for the majority of anthropogenic $CO_2$ release. The **STEP** process, to efficiently recover carbon dioxide and in lieu of these industrial processes, can provide a transition beyond the fossil fuel-electric grid economy.

The top left of Figure 1 includes calculated thermoneutral potentials for $CO_2$ and water splitting reactions. At ambient temperature, the difference between $E_{th}$ and $E_T$ does not indicate an additional heat requirement for electrolysis, as this heat is available via heat exchange with the ambient environment. At ambient temperature, $E_{tn}$ - $E_T$ for $CO_2$ or water is respectively 0.13 and 0.25 V, is calculated (not shown) as 0.15 ±0.1 V for $Al_2O_3$ and $Fe_2O_3$, and 0.28 ±0.3 V for each of the chlorides.

We find that molten electrolytes present several fundamental advantages compared to solid oxides for $CO_2$ electrolysis. (i) Molten carbonate electrolyzer provides $10^3$ to $10^6$ times higher concentration of reactant at the cathode surface than a solid oxide electrolyzer. Solid oxides utilize gas phase reactants, whereas carbonates utilize molten phase reactants. Molten carbonate contains $2 \times 10^{-2}$ mol reducible tetravalent carbon / cm$^3$. The density of reducible tetravalent carbon sites in the gas phase is considerably lower. Air contains 0.03% $CO_2$, equivalent to only $1 \times 10^{-8}$ mol of tetravalent carbon / cm$^3$, and flue gas (typically) contains 10-15% $CO_2$, equivalent to $2 \times 10^{-5}$ mol reducible C(IV) / cm$^3$. Carbonate's higher concentration of active, reducible tetravalent carbon sites, logarithmically decreases the electrolysis potential, and can facilitate charge transfer at low electrolysis potentials. (ii) Molten carbonates can directly absorb atmospheric $CO_2$, whereas solid oxides require an energy consuming pre-concentration process. (iii) Molten carbonates electrolyses are compatible with both solid and gas phase products. (iv) Molten processes have an intrinsic thermal buffer not found in gas phase systems. Sunlight intensity varies over a 24 hour cycle, and more frequently with variations in cloud cover. This disruption to other solar energy conversion processes is not necessary in molten salt processes. For example as discussed in Section 4.3, the thermal buffer capacity of molten salts has been effective for solar to electric power towers to operate 24/7. These towers concentrate solar thermal energy to heat molten salts, which circulate and via heat exchange boil water to drive conventional mechanical turbines.

3. Demonstrated STEP Processes

3.1a STEP Hydrogen

STEP occurs at both higher electrolysis and higher solar conversion efficiencies than conventional room temperature photovoltaic (PV) generation of hydrogen. Experimentally, we demonstrated a sharp decrease in the water splitting potential in an unusual molten sodium hydroxide medium, Figure 3, and as shown in Figure 4, three series connected Si CPVs efficiently driving two series molten hydroxide water splitting cells at 500°C to generate hydrogen.[55,56]

Recently we have considered the economic viability of solar hydrogen fuel production. That study provided evidence that the STEP system is an economically viable solution for the production of hydrogen.[55,56]

3.1b Solar Thermal Electrochemical <u>Pressure</u> Process

The excess (previously unused) sub-bandgap solar energy from solar photovoltaics can be used to increase, not only the temperature,[75] but also the pressure of useful electrosynthetic reactions.[76] This increase in pressure, when applied to exogenic reactions, can also increase the efficiency of solar to chemical energy converstion. For molten electrolyte water splitting, higher temperatures molten will tend to dehydrate the electrolyte decreasing the coulombic efficiency of hydrogen generation.[76] However, sub-band gap and solar thermal energy can be used to continue to decrease the water splitting electrolysis potential by increasing the pressure as illustrated in Figure 5.

3.2 STEP Carbon capture

In this process carbon dioxide is captured directly, without the need to pre-concentrate dilute $CO_2$, using a high temperature electrolysis cell powered by sunlight in a single step. Solar thermal energy decreases the energy required for the endergonic conversion of carbon dioxide and kinetically facilitates electrochemical reduction, while solar visible generates electronic charge to drive the electrolysis. $CO_2$ can be captured as solid carbon and stored, or used as carbon monoxide to feed chemical or synthetic fuel production. Thermodynamic calculations are used to determine, and then demonstrate, a specific low energy, molten carbonate salt pathway for carbon capture.

Prior investigations of the electrochemistry of carbonates in molten salts tended to focus on reactions of interest to fuel cells,[69] rather than the (reverse) electrolysis reactions of relevance to the STEP reduction of carbon dioxide, typically in alkali carbonate mixtures. Such mixtures substantially lower the melting point compared to the pure salts, and would provide the thermodynamic maximum voltage for fuel cells. However, the electrolysis process is maximized in the opposite temperature domain of fuel cells, that is at elevated temperatures which decrease the energy of electrolysis, as schematically delineated in Scheme 1. These conditions provide a new opportunity for effective $CO_2$ capture.

$CO_2$ electrolysis splitting potentials are calculated from the thermodynamic free energy components of the reactants and products[3,4,61b] as $E = -\Delta G(\text{reaction})/nF$, where n= 4 or 2 for the respective conversion of $CO_2$ to the solid carbon or carbon monoxide products. As calculated using the available thermochemical enthalpy and entropy of the starting components, and as summarized in the left side of Figure 6, molten $Li_2CO_3$, via a $Li_2O$ intermediate, provides a preferred, low energy route compared to $Na_2CO_3$ or $K_2CO_3$ (via $Na_2O$ or $K_2O$), for the conversion of $CO_2$. High temperature is advantageous as it decreases the free energy energy necessary to drive the STEP enodthermic process. The carbonates, $Li_2CO_3$, $Na_2CO_3$ and $K_2CO_3$, have respective melting points of 723 °C, 851 °C and 891 °C. Molten $Li_2CO_3$ not only requires lower thermodynamic electrolysis energy, but in addition has higher conductivity (6 S cm$^{-1}$) than that of $Na_2CO_3$ (3 S cm$^{-1}$) or $K_2CO_3$ (2 S cm$^{-1}$) near the melting point.[77] Higher conductivity is desired as it leads to lower electrolysis ohmic losses. Low carbonate melting points are achieved by a eutectic mix of alkali carbonates ($T_{mp}$ $Li_{1.07}Na_{0.93}CO_3$: 499°C; $Li_{0.85}Na_{0.61}K_{0.54}CO_3$: 393°C). Mass transport is also improved at higher temperature; the conductivity increases from 0.9 to 2.1 S cm$^{-1}$ with temperature increase from 650 °C to 875 °C for a 1:1:1 by mass mixture of the three alkali carbonates.[78]

In 2009 we showed that molten carbonate electrolyzers can provide an effective media for solar splitting of $CO_2$ at high conversion efficiency. In 2010 Lubormirsky, et al, and our group separately reported that molten lithiated carbonates provide a particularly effective medium for the electrolytsis reduction of carbon dioxide.[4,79] As we show in the photograph in Figure 6, at 750°C, carbon dioxide is captured in molten lithium carbonate electrolyte as solid carbon by reduction at the cathode at low electrolysis potential. It is seen in the cyclic voltammetry, CV, that a solid carbon peak that is observed at 750°C is not evident at 950°C. At temperatures less than ~900 °C in the molten electrolyte, solid carbon is the preferred $CO_2$ splitting product, while carbon monoxide is the preferred product at higher temperature. As seen in the main portion of the figure, the electrolysis potential is < 1.2V at either 0.1 or 0.5 A/cm$^2$, respectively at 750 or 850°C. Hence, the electrolysis energy required at these elevated, molten temperatures is less than the minimum energy required to split $CO_2$ to CO at 25°C:

$$CO_2 \rightarrow CO + 1/2 O_2 \qquad E°(T=25°C) = 1.33 \text{ V} \qquad [15]$$

The observed experimental carbon capture correlates with:

$$Li_2CO_3(\text{molten}) \rightarrow C(\text{solid}) + Li_2O(\text{dissolved}) + O_2(\text{gas}) \qquad [16A]$$
$$Li_2CO_3(\text{molten}) \rightarrow CO(\text{gas}) + Li_2O(\text{dissolved}) + 1/2 O_2(\text{gas}) \qquad [16B]$$

When $CO_2$ is bubbled in, a rapid reaction back to the original lithium carbonate is strongly favored:

$$Li_2O(\text{dissolved}) + CO_2(\text{gas}) \rightarrow Li_2CO_3(\text{molten}) \qquad [17A]$$
$$Li_2CO_3 \rightleftharpoons Li_2O + CO_2 \qquad [17B]$$

In the presence of carbon dioxide, reaction 17A is strongly favored (exothermic), and the rapid reaction back to the original lithium carbonate occurs while $CO_2$ is bubbled into molten lithium carbonate containing the lithium oxide.

The carbon capture reaction in molten carbonate, combines eqs 16 and 17:

$$CO_2(\text{gas}) \rightarrow C(\text{solid}) + O_2(\text{gas}) \quad T \leq 900°C \qquad [18A]$$
$$CO_2(\text{gas}) \rightarrow CO(\text{gas}) + 1/2 O_2(\text{gas}) \quad T \geq 950°C \qquad [18B]$$

The electrolysis of carbon capture in molten carbonates can occur at lower experimental electrolysis potentials than the unit activity potentials calculated in Figure 6. A constant influx of carbon dioxide to the cell maintains a low concentration of $Li_2O$, in accord with reaction 23. The activity ratio, $\Theta$, of the carbonate reactant to the oxide product in the electrolysis chamber, when high, decreases the cell potentials with the Nernst concentration variation of the potential in accord with eq 16, as:

$E_{CO2/X}(T) = E°_{CO2/X}(T) - 0.0592V \cdot T(K)/(n \cdot 298K) \cdot \log(\Theta)$;

$n=4$ or $2$, for $X= C_{solid}$ or CO product     [19]

For example from eq 19, the expected cell potential at 950°C for the reduction to the CO product is $E_{CO2/CO} = 1.17$ V $-(0.243V /2) \cdot 4= 0.68$ V, with a high $\Theta=10,000$ carbonate/oxide ratio in the electrolysis chamber. As seen in the Figure 6 photo, $CO_2$ is captured in 750°C $Li_2CO_3$ as solid carbon by reduction at the cathode at low electrolysis potential. The carbon formed in the electrolysis in molten $Li_2CO_3$ at 750°C is in quantitative accord with the 4 e-reduction of eq 16A, as determined by (i) mass, at constant 1.25 A for both 0.05 and 0.5 A/cm$^2$ (large and small electrode) electrolyses (the carbon is washed in a sonicator, and dried at 90°C), by (ii) ignition (furnace combustion at 950°C) and by (iii) volumetric analysis in which $KIO_3$ is added to the carbon, converted to $CO_2$ and $I_2$ in hot phosphoric acid ($5C +4KIO_3 +4H_3PO_4 \rightarrow 5CO_2 +2I_2 +2H_2O + 4KH_2PO_4$), the liberated $I_2$ is dissolved in 0.05 M KI and titrated with thiosulfate using a starch indicator. We also observe the transition to the carbon monoxide product with increasing temperature. Specifically, while at 750°C the molar ratio of solid carbon to CO-gas formed is 20:1, at 850° in molten $Li_2CO_3$, the product ratio is a 2:1, at 900°C, the ratio is 0.5:1, and at 950°C the gas is the sole product. Hence, in accord with Figure 2, switching between the C or CO product is temperature programmable.

We have replaced Pt, with Ni, nickel alloys (inconel and monel), Ti and carbon, and each are effective carbon capture cathode materials. Solid carbon deposits on each of these cathodes at similar overpotential in 750°C molten $Li_2CO_3$. For the anode, both platinum and nickel are effective, while titanium corrodes under anodic bias in molten $Li_2CO_3$. As seen in the right side of Figure 6, electrolysis anodic overpotentials in $Li_2CO_3$ electrolysis are comparable, but larger than cathodic overpotentials, and current densities of over 1A cm$^{-2}$ can be sustained. Unlike other fuel cells, carbonate fuel cells are resistant to poisoning effects,[69] and are effective with a wide range of fuels, and this appears to be the same for the case in the reverse mode (to capture carbon, rather than to generate electricity). Molten $Li_2CO_3$ remains transparent and sustains stable electrolysis currents after extended (hours/days) carbon capture over a wide range of electrolysis current densities and temperatures.

As delineated in Section 2.3, in practice, either STEP or Hy-STEP modes are useful for efficient solar carbon capture. $CO_2$ added to the cell is split at 50% solar to chemical energy conversion efficiency by series coupled lithium carbonate electrolysis cells driven at maximum power point by an efficient CPC. Experimentally, we observe the facile reaction of $CO_2$ and $Li_2O$ in molten $Li_2CO_3$. We can also calculate the thermodynamic equilibrium conditions between the species in the system, eq 3B. Using the known thermochemistry of $Li_2O$, $CO_2$ and $Li_2CO_3$,[61b] we calculate the reaction free-energy of eq 1, and from this calculate the thermodynamic equilibrium constant as a function of temperature. From this equilibrium constant, the area above the curve on the left side of Figure 6 presents the wide domain (above the curve) in which $Li_2CO_3$ dominates, that is where excess $CO_2$ reacts with $Li_2O$ such that $p_{CO2} \bullet a_{Li2O} < a_{Li2CO3}$. This is experimentally verified when we dissolve $Li_2O$ in molten $Li_2CO_3$, and inject $CO_2$(gas). Through the measured mass gain, we observe the rapid reaction to $Li_2CO_3$. Hence, $CO_2$ is flowed into a solution of 5% by weight $Li_2O$ in molten $Li_2CO_3$ at 750°C, the rate of mass gain is only limited by the flow rate of $CO_2$ into the cell (using an Omega FMA 5508 mass flow controller) to react one equivalent of $CO_2$ per dissolved $Li_2O$. As seen in the measured thermogravimetric analysis on the right side of Figure 7, the mass loss in time is low in lithium carbonate heated in an open atmosphere (~0.03% $CO_2$) up to 850°C, but accelerates when heated to 950°C. However the 950°C mass loss falls to nearly zero, when heated under pure (1 atm) $CO_2$. Also in accord with eq 1 added

Li$_2$O shifts the equilibrium to the left. As seen in the figure in an open atmosphere, there is no mass loss in a 10% Li$_2$O, 90% Li$_2$CO$_3$ at 850°C, and the Li$_2$O containing electrolyte absorbs CO$_2$ (gains mass) at 750°C to provide for the direct carbon capture of atmospheric CO$_2$, without a CO$_2$ pre-concentration stage. This consists of the absorption of atmospheric CO$_2$ (in molten Li$_2$CO$_3$ containing Li$_2$O, to form Li$_2$CO$_3$), combined with a facile rate of CO$_2$ splitting due to the high carbonate concentration, compared to the atmospheric concentration of CO$_2$, and the continuity of the steady-state of concentration Li$_2$O, as Li$_2$CO$_3$ is electrolyzed in eq 16.

Recently, we have probed the minimum electrolysis energy calculated and observed for the electoltyic splitting of carbon dioxide to solid carbon, investigated barium carbonate composite electrolytes,[80] and as delineated in the upcomng sections extensively studied carbon dioxide electrolysis for solar fuel and carbon nanomaterial synthesis and for mitgation of this greenhouse gas.

3.3 Carbon Nanomaterials

In 2010, as described in Figure 6, we reported on a STEP solid carbon synthesis by the electrolytic splitting of CO$_2$ in molten carbonate.[4] Recently, we discovered that this process not only generates graphite, but more specifically can lead to a high product yield of carbon nanotubes. The electrosynthesis requires low energy,[82] and does not require exotic, nor expensive reagents.[81] Carbon nanotubes as a commodity are valued for their high strength to mass ratio, high thermal and electrical conductivity, high catalytic activity, and battery and nanotechnology applications. Carbon dioxide is the reactant. Hence the transformation provides an incentive to remove and mitigate the greenhouse gas CO$_2$. $^{13}$CO$_2$ tracking demonstrates that atmospheric (or flue gas) CO$_2$ provides the carbon building blocks for the carbon nanotubes, CNTs.[83] We observe that that the (less expensive) natural carbon isotope mix ($^{12}_{0.99}{}^{13}_{0.01}$CO$_2$, rather than $^{13}$CO$_2$) produced the more valuable (carbon nanotube, rather than an alternative nanofiber) product.[83] CNTs are formed at high yield in either pure Li$_2$CO$_3$, and also in mixed Li/Na, and Li/Ba and/or Ca molten carbonate electrolytes.[81,83-87,91] The $CO_2$ synthesized CNTs are a durable, high capacity anode material for Li-ion and Na-ion batteries. Oxide added to the carbonate electrolyte adds $sp^3$ defects, tangling the CNT morphology and further enhancing battery storage behavior.[86] The production of doped CNTs, including boron, phosphorous, nitrogen and sulfur may be accomplished by the direct addition of dopants into the electrolyte used in the synthesis.[84,87,91] Boron doping, through addition of metaborate to the carbonate electrolyte during synthesis, provides a factor of ten enhancement in CNT electrical conductivity.

We have termed the high yield, molten carbonate electrolytic transformation of $CO_2$ to CNTs as the C2CNT process. As illustrated in Scheme 4, the synthesis does not require pre-concentration of $CO_2$, and the direct conversion of both atmospheric and industrial flue-gas concentrations is readily achieved.[81-89,91]

The observed growth mechanism of transition metal nucleiated growth has been modeled by DFT energy minimization (RSC) and modified through experimental variation of the nucleating agent.[81,87,90,91] The mechanism includes tethered growth from the cathode, and can lead to unusually long CNTs, particularly suitable for woven materials, as shown in Figure 8. C2CNT integration into industrial processes, including fossil fuel electric plants and cements, has been probed, to eliminate $CO_2$ emissions, and simultaneously producing a useful, valuable product.[88,89] A wide variety of different CNTs are readily synthesized, including doped doped CNTs grown directly by adding dopants to the electrolyte, thick and thin walled CNTs, and short or long CNTs.[81,83-91]

3.4 Fuels

The electrosynthesis of carbon dioxide (reactions 16A and 16B) in molten carbonates or carbonate/hydroxide mixed electrolytes yields several solar fuels, including $CH_4$, at high solar efficiencies. These fuels include hydrogen and carbon, "sungas" (solar generated syngas, CO

+H$_2$),[75,92,93] and the efficient, simultaneous, direct co-generation of methane and hydrogen Figure 9:[94]

Molten carbonate in electrolyte CO$_2$ splitting: CO$_2$ + 2HO$_2$ → CH$_4$ + 2O$_2$     [20]

Molten hydroxide in electrolyte water splitting: H$_2$O → H$_2$ + 1/2O$_2$     [21]

3.5 STEP Ammonia

Ammonia is a critical resource to produce the world's fertilizer, but H$_2$ production for the synthesis releases carbon dioxide is a substantial contributer to anthropogenic CO$_2$ buildup in the enviroment. The conventional Haber-Bosch ammonia process uses H$_2$ as a reactant, principally produced by natural gas steam reformation (CH$_4$ + 2H$_2$O → 4H$_2$ + CO$_2$). Ammonia production was 1.45x10$^8$ tons in 2014; emitting 2x10$^8$ tonnes of the greenhouse gas CO$_2$. CO$_2$-free reactions without the addition of hydrogen are needed. To this end, we utilized Fe$_2$O$_3$ as an electrocatalyst in the molten hydroxide electrolyte synthesis of ammonia directly from air and water as illustatrated in Figure 10.[95,96] Metallic iron was determined as the chemical intermediate,[96] and the ammonia iron oxide electrocatalyst can be stand-alone,[95,96] or isolated on activated carbon.[97]

    Nanoparticles of Fe$_2$O$_3$ catalyze ammonia generation with air and steam during molten hydroxide electrolysis (76a,b):

Electrochemical:     Fe$_2$O$_3$ → 2Fe + 3/2O$_2$     [22A]

Chemical:     2Fe + N$_2$ + 3H$_2$O → 2NH$_3$ + Fe$_2$O$_3$     [22B]

Net:     N$_2$ + 3H$_2$O → 2NH$_3$ + 3/2O$_2$     [22C]

    Here, we focus on the electrolysis component for STEP ammonia, and unlike the earlier studies the ammonia iron catalyst is formed *in-situ*, rather than by addition. STEP requires high temperature with molten carbonate to deposit and reform the iron catalyst necessary for

sustainable iron. Molten hydroxide can be added to establish a foundation for proton availability (2MOH ⇌ $M_2O$ + $H_2O$). However, high temperature dehydrates the electrolyte. A "goldilocks" intermediate temperature range is established in Figure 10 in which STEP ammonia is sustainable in a mixed molten carbonate/hydroxide electrolyte. High electrolysis currents ensure small, reactive iron particles facilitating higher rates of ammonia generation.

Three molten carbonate electrolytes and temperature domains are explored: a low temperature domain window (< 400°C) opened up by use of a low melting point mixed alkali ($Li_xNa_yK_zCO_3$) carbonate eutectic, a high temperature domain electrolyte based on the higher melting point $Li_2CO_3$ (mp 723°C), and an intermediate alkali earth/ $Li_2CO_3$ mix functional as an electrolyte in the 600°C range. A $Li_{1.6}Ba_{0.3}Ca_{0.1}CO_3$ with 6m LiOH and 1.5m $Fe_2O_3$ electrolyte is molten and not viscous at 650°C. This intermediate temperature mix exhibited the highest and most stable rate of ammonia generation during a 250mA electrolysis in Figure 10. As shown in the figure, even higher ammonia generation rates are observed in this 650 °C electrolyte when the same total electrolysis charge is applied, but pulsed at ten-fold higher current (2.5A) for 1/10 the time, over each 2h cycle (12 minutes 2.5A followed by 108 minutes 0A, repeat cycle) although this configuration passivates rapidly as observed in the sharp tail-off of the observed ammonia generation rate.

3.6 STEP Cement

Cement production accounts for 5-6% of all anthropogenic $CO_2$ emissions. Society consumes over $3 \times 10^{12}$ kg of cement annually, and the cement industry releases 9 kg of $CO_2$ for each 10 kg of cement produced. The majority of $CO_2$ emissions occurs during the decarbonation of limestone ($CaCO_3$) to lime (CaO) described in eq 23A, and the remainder (30 to 40%) from burning fossil fuels, such as coal, to heat the kiln reactors to ~900°C, eq 23B;[98]

$CaCO_3 + Q_{heat} \rightarrow CaO + CO_2$ 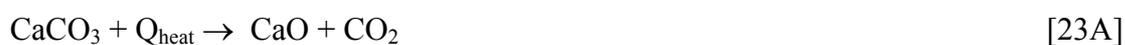 [23A]

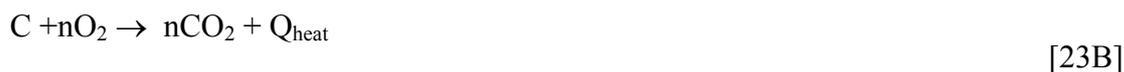

C + nO$_2$ → nCO$_2$ + Q$_{heat}$  [23B]

Instead, STEP cement eliminates CO$_2$ by electrolysis to form carbon:

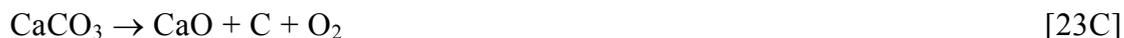

CaCO$_3$ → CaO + C + O$_2$  [23C]

We have investigated two modes of STEP cement as illustrated in Scheme 5. In the direct mode the limestone is dissolved in the molten electrolyte and directly electrolyzed to the desirse calcium oxide (lime) product, as well as solid carbon (and oxygen) gas in lieu of a CO$_2$ emission.[98] In the indirect mode the solid limestone is split by conventional thermal decomposition into the desired calcium oxide product, but instead of escaping into the atmosphere, the emitted carbon dioxide gas (both from reaction 1 and reaction 2) is collected, dissolved in molten carbonate and eliminated by electrolysis to solid carbon and oxygen.

Interestingly, and as shown in Figure 11, we observed the solubilities of lime and limetone, compared to their behavior in water, exhibit opposite solubility in molten carbonate (calcium carbonate is highly soluble, while calicium oxide has low solubility). This has the benefit of allowing the desired lime product to be extracted by precipition in the direct STEP cement mode, while the indirect mode has the ease of simultaneous collection of CO2 emitted from both reactions 1 and 2.[99] As noted in a prior section, the solid carbon product can be valuable carbon nanotubes providing a substantial economic incentive for this carbon mitigation process.

3.7 STEP Organics

Solar heat can also be utilized to facilitate the selectivity and enhance the kinetics of organic reactions such as the reaction of toluene to benzoic acid as illustrated in Scheme 6.[100-104] To date, the organic electrosyntheses probed have been electrocatalytic and exergonic, rather than endergonic, but as with the previous STEP processes substantial solar thermal, kinetic enhancement is demonstrated. STEP organic has been suggested for wastewater treament and can co-generate hydrogen. Most recently, in an effort led by Wang *et al*, these processes have been energized through addition of a photocatalytic electrode (near-UV driven $TiO_2$ enhancement) to the solar thermal and photovoltaic driven electrochemical process, as illustrated in Scheme 7 for the observed reaction of tolune to benzoic acid at an irradiated $TiO_2$.[104]

3.8 STEP Iron

A fundamental change in the understanding of iron oxide thermochemistry can open a facile, new $CO_2$-free route to iron production. Along with control of fire, iron production is one of the founding technological pillars of civilization, but is a major source of $CO_2$ emission. In industry, iron is still produced by the carbothermal greenhouse gas intensive reduction of iron oxide by carbon-coke, and a carbon dioxide free process to form this staple is needed.

The earliest attempt at electrowinning iron (the formation of iron by electrolysis) from carbonate appears to have been in 1944 in the unsuccessful attempt to electrodeposit iron from a sodium carbonate, peroxide, metaborate mix at 450-500°C, which deposited sodium and magnetite (iron oxide), rather than iron.[107,108] Other attempts[108] have focused on iron electrodepostion from molten mixed halide electrolytes, which has not provided a successful route to form iron,[109,110] or aqueous iron electrowinning [111-114] that is hindered by the high thermodynamic potential (E°=1.28 V) and diminished kinetics at low temperature.

We present a novel route to generate iron metal by the electrolysis of dissolved iron oxide salts in molten carbonate electrolytes,[5,6] unexpected due to the reported insolubility of iron oxide in carbonates, optimize the STEP Iron yield, and demonstrate size control of STEP Iron synthesis.[105,106] We report high solubility of lithiated iron oxides, and facile charge transfer that produces the staple iron at high rate and low electrolysis energy, and can be driven by conventional electrical sources, but is also demonstrated with STEP procesess that decreases or eliminates a major global source of greenhouse gas emission.[3,4]

As recently as 1999, the solubility of ferric oxide, $Fe_2O_3$, in 650°C molten carbonate was reported as very low, a $10^{-4.4}$ mole fraction in lithium/potassium carbonate mixtures, and was reported as invariant of the fraction of $Li_2CO_3$ and $K_2CO_3$.[115] Low solubility, of interest to the optimization of molten carbonate fuel cells, had likely discouraged research into the electrowinning of iron metal from ferric oxide in molten lithium carbonate. Rather than the prior part per million reported solubility, we find higher Fe(III) solubilities, on the order of 50% in carbonates at 950°C. The CV of a molten $Fe_2O_3$ $Li_2CO_3$ mixture presented in Figure 12, and exhibits a reduction peak at -0.8 V, on Pt (gold curve); which is more pronounced at an iron electrode (light gold curve). At constant current, iron is clearly deposited. The cooled deposited product contains pure iron metal and trapped salt, and changes to rust color with exposure to water (figure photo). The net electrolysis is the redox reaction of ferric oxide to iron metal and $O_2$, eq 14. The deposit is washed, dried, and is observed to be reflective, grey metallic, responds to an external magnetic field, and consists of dendritic iron crystals.

The two principle natural ores of iron are hematite ($Fe_2O_3$) and the mixed valence $Fe^{2+/3+}$ magnetite ($Fe_3O_4$). We observe that, $Fe_3O_4$ is also highly soluble in molten $Li_2CO_3$, and may also be reduced to iron with the net electrolysis reaction:

$Fe_3O_4 \rightarrow 3Fe + 2O_2$ 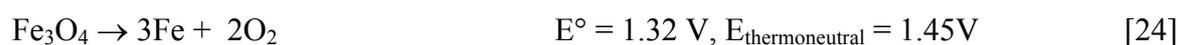 $E° = 1.32$ V, $E_{thermoneutral} = 1.45$V [24]

$Fe_3O_4$ electrolysis potentials run parallel, but ~0.06 V higher, than those of $Fe_2O_3$ in Figure 1. The processes are each endergonic ; the required electrolysis potential decreases with increasing temperature. For $Fe_3O_4$ in Figure 12, unlike the single peak evident for $Fe_2O_3$, two reduction peaks appear in the CV at 800°C. Following the initial cathodic sweep (indicated by the left arrow), the CV exhibits two reduction peaks, again more pronounced at an iron electrode (grey curve), which appear to be consistent with the respective reductions of $Fe^{2+}$ and $Fe^{3+}$. In either $Fe_2O_3$, or $Fe_3O_4$, the reduction occurs at a potential before we observe any reduction of the molten $Li_2CO_3$ electrolyte, and at constant current, iron is deposited. Following 1 hour of electrolysis at either 200 or 20 mA/cm$^2$ of iron deposition, as seen in the Figure 12 photographs, and as with the $Fe_2O_3$ case, the extracted cooled electrode, following extended electrolysis and iron formation, contains trapped electrolyte. Following washing, the product weight is consistent with the eight electron per $Fe_3O_4$ coulombic reduction to iron.

The solid products of the solid reaction of $Fe_2O_3$ and $Li_2CO_3$ had been characterized.[116,117] We prepare and probe the solubility of lithiated iron oxide salts in molten carbonates, and report high Fe(III) solubilities, on the order of 50% in molten carbonates, are achieved via the reaction of $Li_2O$ with $Fe_2O_3$, yielding an effective method for $CO_2$ free iron production.

Lithium oxide, as well as $Fe_2O_3$ or $Fe_3O_4$, each have melting points above 1460°C. $Li_2O$ dissolves in 400-1000°C molten carbonates. We find the solubility of $Li_2O$ in molten $Li_2CO_3$ increases from 9 to 14 m from 750° to 950°C. Following preparation of specific iron oxide salts, we add them to molten alkali carbonate. The resultant Fe(III) solubility is similar when either $LiFeO_2$, or $LiFeO_2$ as $Fe_2O_3 + Li_2O$, is added to the $Li_2CO_3$. As seen in the left side of Figure 13, the solubility of $LiFeO_2$ is over 12 m above 900C° in $Li_2CO_3$.

Solid reaction of $Fe_2O_3$ and $Na_2CO_3$ produces both $NaFeO_2$ and $NaFe_5O_8$ products.[118] As seen in Figure 13, our unlike $Li_2CO_3$, measurements in either molten $Na_2CO_3$ or $K_2CO_3$, exhibit << 1 wt % iron oxide solubility, even at 950°C. However the solubility of ($Li_2O$ + $Fe_2O_3$) is high in the alkali carbonate eutectic, $Li_{0.87}Na_{0.63}K_{0.50}CO_3$, and is approximately proportional to the Li fraction in the pure $Li_2CO_3$ electrolyte. Solubility of this lithiated ferric oxide in the $Li_xNa_yK_zCO_3$ mixes provides an alternative molten media for iron production, which compared to pure lithium carbonate, has the disadvantage of lower conductivity,[5] but the advantage of even greater availability, and a wider operating temperature domain range (extending several hundred degrees lower than the pure lithium system).

$Fe_2O_3$ or $LiFe_5O_8$ dissolves rapidly in molten $Li_2CO_3$, but reacts with the molten carbonate as evident in a mass loss, which evolves one equivalent of $CO_2$ per $Fe_2O_3$, to form a steady state concentration of $LiFeO_2$ in accord with the reaction of eq 25A (but occurring in molten carbonate).[6] However, 1 equivalent of $Li_2O$ and 1 equivalent of $Fe_2O_3$, or $LiFeO_2$, dissolves without the reactive formation of $CO_2$. This is significant for the electrolysis of $Fe_2O_3$ in molten carbonate. As $LiFeO_2$ is reduced $Li_2O$ is released, eq 25B, facilitating the continued dissolution of $Fe_2O_3$ without $CO_2$ release or change in the electrolyte, More concisely, iron production via hematite in $Li_2CO_3$ is given by I and II:

**I** dissolution in molten carbonate   $Fe_2O_3 + Li_2O \rightarrow 2LiFeO_2$      [25A]

**II** electrolysis, $Li_2O$ regeneration: $2LiFeO_2 \rightarrow 2Fe + Li_2O + 3/2O_2$      [25B]

**Iron Production**, $Li_2O$ unchanged (I+II): $Fe_2O_3 \rightarrow 2Fe + 3/2O_2$      [26]

As indicated in Figure 12, a molar excess, of greater than 1:1 of $Li_2O$ to $Fe_2O_3$ in molten $Li_2CO_3$, will further inhibit the eq 1 disproportionation of lithium carbonate. The right side of Figure 13 summarizes the thermochemical calculated potentials constraining iron production in molten carbonate. Thermodynamically it is seen that at higher potential, steel (iron containing carbon) may be directly formed via the concurrent reduction of $CO_2$, which we observe in the $Li_2CO_3$ at higher electrolysis potential, as $Li_2CO_3 \rightarrow C + Li_2O + O_2$, followed by carbonate regeneration via eq 3, to yield by electrolysis in molten carbonate:

**Steel Production**:   $Fe_2O_3 + 2xCO_2 \rightarrow 2FeC_x + (3/2+2x)O_2$      [27]

From the kinetic perspective, a higher concentration of dissolved iron oxide improves mass transport, decreases the cathode overpotential and permits higher steady-state current densities of iron production, and will also substantially decrease the thermodynamic energy needed for the reduction to iron metal. In the electrolyte Fe(III) originates from dissolved ferric oxides, such as $LiFeO_2$ or $LiFe_5O_8$. The potential for the $3e^-$ reduction to iron varies in accord with the general Nerstian expression, for a concentration [Fe(III)], at activity coefficient, $\alpha$:

   $E_{Fe(III/0)} = E°_{Fe(III/0)} + (RT/nF) \log(\alpha_{Fe(III)} [Fe(III)])^{1/3}$      [28]

This decrease in electrolysis potential is accentuated by high temperature and is a ~0.1 V per decade increase in Fe(III) concentration at 950°C. Higher activity coefficient, $\alpha_{Fe(III)} > 1$, would further decrease the thermodynamic potential to produce iron. The measured electrolysis potential is presented on the right of Figure 12 for dissolved Fe(III) in molten $Li_2CO_3$, and is low. For example 0.8V sustains a current density of 500 mA cm$^{-2}$ in 14 m Fe(III) in $Li_2CO_3$ at 950°C. Higher temperature, and higher concentration, lowers the electrolysis voltage, which can be considerably less than the room potential required to convert $Fe_2O_3$ to iron and oxygen. When an external source of heat, such as solar thermal, is available then the energy savings over room temperature iron electrolysis are considerable.

Electrolyte stability is regulated through control of the $CO_2$ pressure and/or by dissolution of excess $Li_2O$. Electrolyte mass change was measured in 7 m $LiFeO_2$ & 3.5 m $Li_2O$ in molten $Li_2CO_3$ after 5 hours. Under argon there is a 1, 5 or 7 wt% loss respectively at 750°C, 850°C or 950°C), through $CO_2$ evolution. Little loss occurs under air (0.03% $CO_2$), while under pure $CO_2$ the electrolyte gains 2-3 wt% (external $CO_2$ reacts with dissolved $Li_2O$ to form $Li_2CO_3$).

The endergonic nature of the new synthesis route, that is the decrease in iron electrolysis potential with increasing temperature, provides a low free energy opportunity for the STEP process. In this process, solar thermal provides heat to decrease the iron electrolysis potential, Figure 12, and solar visible generates electronic charge to drive the electrolysis. A low energy route for the carbon dioxide free formation of iron metal from iron ores is accomplished by the synergistic use of both visible and infrared sunlight. This provides high solar energy conversion efficiencies, Figure 2, when applied to eqs 14, 24 and 26 in a molten carbonate electrolytes. We again use a 37% solar energy conversion efficient concentrator photovoltaic (CPV) as a convenient power source to drive the low electrolysis energy iron deposition without $CO_2$ formation in $Li_2CO_3$,[3] as schematically represented in Figure 14.

A solar/wind hybrid Solar Thermal Electrochemical Production iron electrolysis process is also demonstrated.[6] In lieu of solar electric, electronic energy can be provided by alternative renewables, such as wind. As shown on the right side of Figure 14, in this Hy-STEP example, the electronic energy is driven by a wind turbine and concentrated sunlight is only used to provide heat to decrease the energy required for iron splitting. In this process, sunlight is concentrated to provide effective heating, but is not split into separate spectral regions as in our alternative implmentation. Hy-STEP iron production is measured with a 31.5"x44.5" Fresnel lens (Edmund Optics) which concentrates sunlight to provide temperatures of over 950°C, and a Sunforce-44444 400 W wind turbine provides electronic charge, charging series nickel metal hydride, MH, cells at 1.5V). Each MH cell, provides a constant discharge potential of 1.0-1.3 V, which are each used to drive one or two series connected iron electrolysis cells as indicated in the right side of Figure 14, containing 14 m Fe(III) molten $Li_2CO_3$ electrolysis cells. Electrolysis current is included in the lower right of Figure 14. Iron metal is produced. Steel (iron containing carbon) may be directly formed via the concurrent reduction of $CO_2$, as will be delineated in an expanded study.

3.9 STEP Chlorine and magnesium production (chloride electrolysis).

The predominant salts in seawater (global average 3.5±0.4% dissolved salt by mass) are NaCl (0.5 M) and $MgCl_2$ (0.05 M). The electrolysis potential for the industrial chlor-alkali reaction exhibits little variation with temperature, and hence the conventional generation of chlorine by electrolysis, eq 11, would not benefit from the inclusion of solar heating.[3] However, when confined to anhydrous chloride splitting, as exemplified in the lower portion of Figure 1, the calculated potential for the anhydrous electrolysis of chloride salts is endergonic for the electrolyses, which generate a chlorine and metal product. The application of excess heat, as through the **STEP** process, decreases the energy of electrolysis and can improve the kinetics of charge tranfer for the eq 12 range of chloride splitting processes. The thermodynamic electrolysis potential for the conversion of NaCl to sodium and chlorine decreases, from 3.24V at the 801°C melting point, to 2.99 V at 1027°C.[3] Experimentally, at 850°C in molten NaCl, we observe the expected, sustained generation of yellow-green chlorine gas at a platinum anode and of liquid sodium (mp 98 °C) at the cathode.

Electrolysis of a second chloride salt, $MgCl_2$, is also of particular interest. The magnesium, as well as the chlorine, electrolysis products are significant societal commodities. Magnesium metal, the third most commonly used metal, is generally produced by the reduction of calcium magnesium carbonates by ferrosilicons at high temperature,[119] which releases substantial levels of carbon dioxide contributing to the anthropogenic greenhouse effect. However, traditionally, magnesium has also been produced by the electrolysis of magnesium chloride, using steel cathodes and graphite anodes, and alternative materials have been invesitgated.[120]

Of significance, here to the **STEP** process, is the highly endergonic nature of anhydrous chloride electrolysis, such as for $MgCl_2$ electrolysis, in which solar heat will also decrease the energy (voltage) needed for the electrolysis. The rest potential for electrolysis of magnesium chloride decreases from 3.1 V, at room temperature, to 2.5 V at the 714°C melting point. As seen in Figure 15, the calculated thermodynamic potential for the electrolysis of magnesium chloride continues to decrease with increasing temperature, to ~2.3 V at 1000 °C. The 3.1 V energy stored in the magnesium and chlorine room temperature products, when formed at 2.3 V, provide an energy savings of 35%, if sufficient heat applied to the process can sustain this lower formation potential. Figure 15 also includes the experimental decrease in the $MgCl_2$ electrolysis potential with increasing temperature in the lower right portion. In the top portion of the figure, the concurrent shift in the cyclic voltammogram is evident, decreasing the potential peak of magnesium formation, with increasing temperature from 750°C to 950°C. Sustained electrolysis and generation of chlorine at the anode and magnesium at the cathode (Figure 15, photo inset) is evident at platinum electrodes. The measured potential during constant current electrolysis at 750°C in molten $MgCl_2$ at the electrodes is included in the figure.

In the magnesium chloride electrolysis cell, nickel electrodes yield similar results to platinum, and can readily be used to form larger electrodes. The nickel anode sustains extended chlorine evolution without evident deterioration; the nickel cathode may slowly alloy with deposited magnesium. The magnesium product forms both as the solid and liquid (Mg mp 649 °C). The liquid magnesium is less dense than the electrolyte, floats upwards, and eventually needs to be separated and removed to prevent an inter-electrode short, or to prevent a reaction with chlorine that is evolved at the anode. In a scaled up cell configuration (not shown in Figure 15, a larger Ni cathode (200 cm$^2$ cylindrical nickel sheet (McMaster 9707K35) was employed, sandwiched between two coupled cylindrical Ni sheet anodes (total 200 cm$^2$, of area across from the cathode) in a 250 ml alumina (Adavalue) crucible, and sustains multi- ampere currents. The potential at constant current is initially stable, but this cell configuration leads to electrical shorts, unless liquid magnesium is removed.

One salt source for the STEP generation of magnesium and chlorine from $MgCl_2$ are via chlorides extracted from salt water, with the added advantage of the generation of less saline water as a secondary product. In the absence of effective heat exchanger, concentrator photovoltaics heat up to over 100°C, which decreases cell performance. Heat exchange with the (non-illuminated side of) concentrator photovoltaics can vaporize seawater for desalinization and simultaneously prevent overheating of the CPV. The simple concentrator STEP mode (coupling super-bandgap electronic charge with solar thermal heat) is applicable when sunlight is sufficient to both generate electronic current for electrolysis and sustain the electrolysis temperature. In cases, requiring both the separation of salts from aqueous solution followed by molten electrolysis of the salts, a single source of concentrated sunlight can be insufficient, to both drive water desalinization, and to also heat and drive electrolysis of the molten salts. Figure 10 includes a schematic representation of a Hybrid-Solar Thermal Electrochemical Production process with separate (i) solar thermal and (ii) photovoltaic field to drive both desalinization and the endergonic carbon dioxide-free electrolysis of the separated salts, or water splitting, to useful products. As illustrated, the separate thermal and electronic sources may each be driven by insolation, or alternatively, can be (i) solar thermal and (ii) (not illustrated) wind, water, nuclear or geothermal driven electronic transfer.

**4. STEP Constraints**

4.1 STEP Limiting Equations

As illustrated on the left side of Scheme 2, the ideal **STEP** electrolysis potential incorporates not only the enthalpy needed to heat the reactants to $T_{STEP}$ from $T_{ambient}$, but also the heat recovered via heat exchange of the products with the inflowing reactant. In this derivation it is convenient to describe this combined heat in units of voltage via the conversion factor nF:

$Q_T \equiv \sum_i H_i(R_i, T_{STEP}) - \sum_i H_i(R_i, T_{ambient}) - \sum_i H_i(C_i, T_{STEP}) + \sum_i H_i(C_i, T_{ambient})$;

$$E_Q(V) = -Q_T(J/mol)/nF \qquad [29]$$

The energy for the process, incorporates $E_T$, $E_Q$, and the non-unit activities, via inclusion of eq 29 into eq 4, and is termed the **STEP** potential, $E_{STEP}$:

$E_{STEP}(T,a) = [-\Delta G°(T) - Q_T - RT \cdot \ln( \prod_{i=1 \text{ to } x} a(R_i)^{r_i} / \prod_{i=1 \text{ to } y} a(P_i)^{p_i} )]/nF$;

$$E°_{STEP}(a=1) = E_T° + E_Q \qquad [30]$$

In a pragmatic electrolysis system, product(s) can be be drawn off at activities that are less than that of the reactant(s). This leads to large activity effects in eq 30 at higher temperature,[3-6,8,53-56] as the RT/nF potential slope increases with T (e.g. increasing 3-fold from 0.0592V/n at 25°C to 0.183V/n at 650°C).

The **STEP** factor, $A_{STEP}$ is the extent of improvement in carrying out a solar driven electrolysis process at $T_{STEP}$, rather than at $T_{ambient}$. For example, when applying the same solar energy, to electronically drive the electrochemical splitting of a molecule which requires only two thirds the electrolysis potential at a higher temperature, then $A_{STEP} = (2/3)^{-1} = 1.5$. In general, the factor is given by:

$$A_{STEP} = E_{STEP}(T_{ambient}, a)/E_{STEP}(T_{STEP}, a); \; e.g. \; T_{ambient}=298K \qquad [31]$$

The **STEP** solar efficiency, $\eta_{STEP}$, is constrained by both photovoltaic and electrolysis conversion efficiencies, $\eta_{PV}$ and $\eta_{electrolysis}$, and the **STEP** factor. In the operational process, passage of electrolysis current requires an additional, combined (anodic and cathodic) overpotential above the thermodynamic potential; that is $V_{redox} = (1+z)E_{redox}$, Mobility and kinetics improve at higher temperature and $\xi(T > T_{ambient}) < \xi(T_{ambient},)$.[65,69] Hence, a lower limit of $\eta_{STEP}(V_T)$ is given by $\eta_{STEP\text{-}ideal}(E_T)$. At $T_{ambient}$, $A_{STEP} = 1$, yielding $\eta_{STEP}(T_{ambient}) = \eta_{PV} \cdot \eta_{electrolysis}$. $\eta_{STEP}$ is additionally limited by entropy and black body constraints on maximum solar energy conversion efficiency. Consideration of a black body source emitted at the sun's surface temperature and collected at ambient earth temperature, limits solar conversion to 0.933 when radiative losses are considered,[121] which is further limited to $\eta_{PV} < \eta_{limit} = 0.868$ when the entropy limits of perfect energy conversion are included.[122] These constraints on $\eta_{STEP\text{-}ideal}$ and the maximum value of solar conversion, are imposed to yield the solar chemical conversion efficiency, $\eta_{STEP}$:

$$\eta_{STEP\text{-}ideal}(T,a) = \eta_{PV} \cdot \eta_{electrolysis} \cdot A_{STEP}(T,a)$$

$$\eta_{STEP}(T,a) \cong \eta_{PV} \cdot \eta_{electrolysis}(T_{ambient},a) \cdot A_{STEP}(T,a); \qquad (\eta_{STEP} < 0.868) \qquad [32]$$

As calculated from eq 3 and the thermochemical component data[61b] and as presented in Figure 1, the electrochemical driving force for a variety of chemicals of widespread use by society, including aluminium, iron, magnesium and chlorine, significantly decreases with increasing temperature.

4.2 Predicted STEP Efficiencies for Solar Splitting of $CO_2$

The global community is increasingly aware of the climate consequences of elevated greenhouse gases. A solution to rising carbon dioxide levels is needed, yet carbon dioxide is a highly stable, noncombustible molecule, and its thermodynamic stability makes its activation energy demanding and challenging. The most challenging stage in converting $CO_2$ to useful products and fuels is the initial activation of $CO_2$, for which energy is required. It is obvious that using traditional fossil fuels as the energy source would completely defeat the goal of mitigating greenhouse gases. A preferred route is to recycle and reuse the $CO_2$ and provide a useful carbon resource. We limit the non-unit activity examples of $CO_2$ mitigation in eq 15 to the case when CO and $O_2$ are present as electrolysis products, which yields $a_{O_2} = 0.5 a_{CO}$, and upon substitution into eq 30:

$$E_{STEP}(T,a) = E°_{STEP}(T) - (RT/2F) \cdot \ln(N); \quad E°(25°C) = 1.333V; \quad N = \sqrt{2}\, a_{CO_2}\, a_{CO}^{-3/2} \qquad [33]$$

The example of $E_{STEP}(T, a \neq 1)$ on the left side of Figure 16 is derived when N= 100, and results in a substantial drop in the energy to split $CO_2$ due to the discussed influence of RT/2F. Note at high temperature conditions in the figure, $E_{STEP} < 0$ occurs, denoting the state in which the reactants are spontaneously formed (without an applied potential). This could lead to the direct thermochemical generation of products, but imposes substantial experimental challenges. To date, analogous direct water splitting attempts, are highly inefficient due to the twin challenges of high temperature material constraints and the difficulty in product separation to prevent back reaction upon cooling.[123] The STEP process avoids this back reaction through the separation of products, which spontaneously occurs in the electrochemical, rather than chemical, generation of products at separate anode and cathode electrodes.

The differential heat required for $CO_2$ splitting, $E_Q$, and the potential at unit activity, $E°_{STEP}$, are calculated and presented in the top of Figure 16. $E_Q$ has also been calculated and is included. $E_Q$ is small (comprising tens of millivolts or less) over the entire temperature range. Hence from eq 30, $E°_{STEP}$ does not differ significantly from the values presented for $E_T°$ for $CO_2$ in Figure 2. $E_{CO2split}(25°C)$ yields $A_{STEP}(T) = 1.333V / E°_{STEP}(T)$ with unit activity, and $A_{STEP}(T) = 1.197V / E_{STEP}(T)$ for the N=100 case. Large resultant **STEP** factors are evident in the left of Figure 16. This generates substantial values of solar to chemical energy conversion efficiency for the **STEP** $CO_2$ splitting to CO and $O_2$.

A **STEP** process operating in the $\eta_{PV} \cdot \eta_{electrolysis}$ range of 0.20 to 0.40 includes the range of contemporary 25 to 45% efficient concentrator photovoltaics,[69] and electrolysis efficiency range of 80 to 90%. From these, the $CO_2$ solar splitting efficiencies are derived from eqs 32 and 33, and are summarized on the right side of **Figure 16**. The small values of $E_{STEP}(T)$ at higher T, generate large **STEP** factors, and result in high solar to chemical energy conversion efficiencies for the splitting of $CO_2$ to CO and $O_2$. As one intermediate example from eq 33, we take the case of an electrolysis efficiency of 80% and a 34% efficient photovoltaic ($\eta_{PV} \cdot \eta_{electrolysis}$ = 0.272). This will drive **STEP** solar $CO_2$ splitting at molten carbonate temperatures (650°C) at a solar conversion efficiency of 35% in the unit activity case, and at 50% when N=100 (the case of a cell with 1 bar of $CO_2$ and ~58 mbar CO).

4.3 Scaleability of STEP Processes

STEP can be used to remove and convert carbon dioxide. As with water splitting, the electrolysis potential required for $CO_2$ splitting falls rapidly with increasing temperature (Figure 1), and we have shown here (Figure 2) that a photovoltaic, converting solar to electronic energy at 37% efficiency and 2.7V, may be used to drive three $CO_2$ splitting, lithium carbonate electrolysis cells, each operating at 0.9V, and each generating a 2 electron CO product. The energy of the CO product is 1.3V (eq 1), even though generated by electrolysis at only 0.9V due to synergistic use of solar thermal energy. As seen in Figure 5, at lower temperature (750°C, rather than 950°C), carbon, rather than CO, is the preferred product, and this 4 electron reduction approaches 100% Faradaic efficiency.

The $CO_2$ STEP process consists of solar driven and solar thermal assisted $CO_2$ electrolysis. Industrial environments provide opportunities to further enhance efficiencies; for example fossil-fueled burner exhaust provides a source of relatively concentrated, hot $CO_2$. The product carbon may be stored or used, and the higher temperature product carbon monoxide can be used to form a myriad of industrially relevant products including conversion to hydrocarbon fuels with hydrogen (which is generated by STEP water splitting in Section 3.1), such as smaller alkanes, dimethyl ether, or the Fischer Tropsch generated middle-distillate range fuels of C11-C18 hydrocarbons including synthetic jet, kerosene and diesel fuels.[124] Both STEP and Hy-STEP represent new solar energy conversion processes to produce energetic molecules. Individual components used in the process are rapidly maturing technologies including wind electric,[125] molten carbonate fuel cells,[69] and solar thermal technologies.[126-131]

It is of interest whether material resources are sufficient to expand the process to substantially impact (decrease) atmospheric levels of carbon dioxide. The buildup of atmospheric $CO_2$ levels from a 280 to 392 ppm occurring over the industrial revolution comprises an increase of $1.9 \times 10^{16}$ mole ($8.2 \times 10^{11}$ metric tons) of $CO_2$,[132] and will take a comparable effort to remove. It would be preferable if this effort results in useable, rather than sequestered, resources. We calculate below a scaled up STEP capture process can remove and convert all excess atmospheric $CO_2$ to carbon.

In STEP, 6 kWh m$^{-2}$ of sunlight per day, at 500 suns on 1 m$^2$ of 38% efficient CPV, will generate 420 kAh at 2.7 V to drive three series connected molten carbonate electrolysis cells to CO, or two series connected series connected molten carbonate electrolysis cells to form solid carbon. This will capture $7.8 \times 10^3$ moles of $CO_2$ day$^{-1}$ to form solid carbon (based on 420 kAh · 2 series cells / 4 Faraday mol$^{-1}$ $CO_2$). The $CO_2$ consumed per day is three fold higher to form the carbon monoxide product (based on 3 series cells and 2 F mol$^{-1}$ $CO_2$) in lieu of solid carbon. The material resources to decrease atmospheric carbon dioxide concentrations with STEP carbon capture, appear to be reasonable. From the daily conversion rate of $7.8 \times 10^3$ moles of $CO_2$ per square meter of CPV, the capture process, scaled to 700 km$^2$ of CPV operating for 10 years can remove and convert all the increase of 1.9 x $10^{16}$ mole of atmospheric $CO_2$ to solid carbon. A larger current density at the electrolysis electrodes, will increase the required voltage and would increase the required area of CPVs. While the STEP product (chemicals, rather than electricity) is different than contemporary concentrated solar power (CSP) systems, components including a tracker for effective solar concentration are similar (although an electrochemical reactor, replaces the mechanical turbine). A variety of CSP installations, which include molten salt heat storage, are being commercialized, and costs are decreasing. STEP provides higher solar energy conversion efficiencies than CSP, and secondary losses can be lower (for example, there are no grid-related transmission losses). Contemporary concentrators, such as based on plastic Fresnel or flat mirror technologies, are relatively inexpensive, but may become a growing fraction of cost as concentration increases.[133] A greater degree of solar concentration, for example 2000 suns, rather than 500 suns, will proportionally decrease the quantity of required CPV to 175 $km^2$, while the concentrator area will remain the same at 350,000 $km^2$, equivalent to 4% of the area of the Sahara desert (which averages ~6 kWh $m^{-2}$ of sunlight per day), to remove anthropogenic carbon dioxide in ten years.

A related resource question is whether there is sufficient lithium carbonate, as an electrolyte of choice for the STEP carbon capture process, to decrease atmospheric levels of carbon dioxide. 700 $km^2$ of CPV plant will generate 5x$10^{13}$ A of electrolysis current, and require ~2 million metric tonnes of lithium carbonate, as calculated from a 2 kg/l density of lithium carbonate, and assuming that improved, rather than flat, morphology electrodes will operate at 5 A/$cm^2$ (1,000 $km^2$) in a cell of 1 mm thick. Thicker, or lower current density, cells will require proportionally more lithium carbonate. Fifty, rather than ten, years to return the atmosphere to pre-industrial carbon dioxide levels will require proportionally less lithium carbonate. These values are viable within the current production of lithium carbonate. Lithium carbonate availability as a global resource has been under recent scrutiny to meet the growing lithium battery market. It has been estimated that the current global annual production of 0.13 million tonnes of LCE (lithium carbonate equivalents) will increase to 0.24 million tonnes by 2015.[133] Potassium carbonate is substantially more available, but as noted in the main portion of the paper can require higher carbon capture electrolysis potentials than lithium carbonate.

**5. Conclusions**

To ameliorate the consequences of rising atmospheric carbon dioxide levels and its effect on global climate change, there is a drive to replace conventional fossil fuel driven electrical production by renewable energy driven electrical production. In addition to the replacement of the fossil fuel economy by a renewable electrical economy, we suggest that a renewable chemical economy is also warranted. Solar energy can be efficiently used, as demonstrated with the STEP process, to directly, and efficiently form the chemicals needed by society without carbon dioxide emission. Iron, a basic commodity, currently accounts for the release of one quarter of worldwide $CO_2$ emissions by industry, which may be eliminated by replacement with the STEP iron process. The unexpected solubility of iron oxides in lithium carbonate electrolytes, coupled with facile charge transfer and a sharp decrease in iron electrolysis potentials with increasing temperature, provides a new route for iron production. Iron is formed without an extensive release of $CO_2$ in a process compatible with the predominant naturally occurring iron oxide ores, hematite, $Fe_2O_3$, and magnetite, $Fe_3O_4$. STEP can also be used in direct carbon capture, and the efficient solar generation of hydrogen and other fuels.

In addition to the removal of $CO_2$, the STEP process is shown to be consistent with the efficient solar generation of a variety of metals, as well as chlorine via endergonic electrolyses. Commodity production and fuel consumption processes are responsible for the majority of industry based $CO_2$ release, and their replacement by STEP processes provide a path to end the root cause of anthropogenic global warming, as a transition beyond the fossil fuel, electrical or hydrogen economy, to a renewable chemical economy based on the direct formulation of the materials needed by society. An expanded understanding of electrocatalysis and materials will advance the efficient electrolysis of STEP's growing porfolio of energetic products.

**Acknowledgements**

The author is grateful to Baohui Wang, Baochen Cui and Hongun Wu for contributions to several of the STEP processe, and for support by US NSF grant 1505830.

1. *(a) On Solar Hydrogen & Nanotechnology*, (Ed: L. Vayssieres), John Wiley and Sons, Weinheim, Germany (2009); (b) *The Solar Generation of Hydrogen: Towards a Renewable Energy Future* (Eds: K. Rajeshwar, S. Licht, R. McConnell), Springer, New York, USA (2008).
2. S. Licht, *Adv. Mat.*, **23**, 5592 (2011).
3. S. Licht, *J. Phys. Chem., C*, **113**, 16283 (2009).
4. S. Licht. B. Wang, S. Ghosh, H. Ayub, D. Jiang, J. Ganely, *J. Phys. Chem. Lett.*, **1**, 2363 (2010).
5. S. Licht, B. Wang, *Chem. Comm.*, **46**, 7004 (2010).
6. S. Licht, H. Wu, Z. Zhang, H. Ayub, *Chem. Comm.*, **47**, 3081 (2011).
7. S. Licht, O. Chityat, H Bergmann, A. Dick, S. Ghosh, H. Ayub, *Int. J. Hyd. Energy*, **35**, 10867 (2010).
8. S. Licht, B. Wang, H. Wu, *J. Phys. Chem., C*, **115**, 11803 (2011).
9. G. Ohla, P. Surya, S. Licht, N. Jackson, *Reversing Global Warming: Chemical Recycling and Utilization of $CO_2$*. Report of the National Science Foundation sponsored 7-2008 Workshop, 17 pages (2009); full report available at: http://www.usc.edu/dept/chemistry/loker/ReversingGlobalWarming.pdf
10. C. Graves, S. Ebbsen, M. Mogensen, K. Lackner, *Renewable Sustainble Energy Rev.*, **15**, 1 (2010).
11. J. Barber, *Chem. Soc. Rev.*, **38**, 185 (2009).
12. A. Stamatiou, P. G. Loutzenhiser, A. Steinfeld, *Energy Fuels*, **24**, 2716 (2010).
13. S. Abanades, M. Chambon, *Energy Fuels*, **24**, 6677 (2010).
14. L. J. Venstrom, J. H. Davidson, *J. Solar Energy Eng. Chem.*, **133**, 011017-1 (2010).

15. W. Chueh, S. Haile, *Phil. Trans. Roy. Soc. A*, **368**, 3269 (2010).
16. J. Miller, M. Allendorf, R. Diver, L. Evans, N. Siegel, J. Stueker, *J. Mat. Sci.*, **43**, 4714 (2008).
17. S. Licht, *Nature*, **330,** 148 (1987).
18. S. Licht, D. Peramunage, *Nature*, **345**, 330 (1990).
19. B. Oregan, M. Gratzel, M. *Nature*, **353**, 737 (1991).
20. S. Licht, *J. Phys. Chem.*, **90**, 1096 (1998).
21. *Semiconductor Electrodes and Photoelectrochemistry*, (Ed: S. Licht), Wiley-VCH, Weinheim, Germany (2002).
22. S. Licht, G. Hodes, R. Tenne, J. Manassen, *Nature*, **326**, 863 (1987).
23. S. Licht, B. Wang, B.; et. al, *Appl. Phys. Lett.*, **74**, 4055 (1999).
24. S. Yan, L. Wan, Z. Li, Z. Zou, *Chem. Comm.*, **47**, 5632 (2011).
25. H. Zhou, T. Fan, D. Zhang, *ChemCatChem,* **3**, 513 (2011).
26. R. Huchinson, E. Holland, B. Carpenter, *Nature Chem.*, **3**, 301 (2011).
27. E. E. Barton, D. M. Rampulla, and A. B. Bocarsly, *J. Am. Chem. Soc.*, **130**, 6342 (2008).
28. S. Kaneco, H. Katsumata, T. Suzuki, K. Ohta, *Chem Eng. J.,*, **92**, 363 (2006).
29. P. Pan, Y. Chen, *Catal, Comm.* **8**, 1546 (2007).
30. A. B. Murphy, *Solar Energy Mat*, *116*, 227 (2008).
31. A. Currao, *Chimia*, **61**, 815 (2007).
32. a) S. R. Narayanan, B. Haines, J. Soler, T. I. Valdez, *J. Electrochem. Soc.*, *158*, A167 (2011), b) C. Delacourt, J. Newman, *ibid.*, *157*, B1911 (2010).
33. E. Dufek, T. Lister, M. McIlwain, *J. Appl. Electrochem.*, *41*, 623 (2011).
34. M. Gangeri, S. Perathoner, S. Caudo, G. Centi, J. Amadou, D. Begin, C. Pham-Huu, M. Ledoux, J. Tessonnier, D. Su, R. Schlogl, *Catalysis Today*, **143**, 4714 (2009).
35. B. Innocent, D. Liaigre, D. Pasquier, F. Ropital, J. Leger, K. Kokoh, *J. Appl. Electrochem.*, *39*, 227 (2009).
36. A. Wang, W. Liu, S. Cheng, D. Xing, J. Zhou, B. Logan, *Intl. J. Hydrogen Energy*, **39**, 3653 (2009).
37. N. Dong-fang, X. Cheng-tian, L. Yi-wen, Z. Li, L. Jiz-xing, *Chem. Res. Chinese U.*, **34**, 708 (2009).
38. D. Chu, G. Qin, X. Yuan, M. Xu, P. Zheng, J. Lu, *ChemSusChem*, *1*, 205 (2008).
39. J. Yano, T. Morita, K. Shimano, Y. Nanmi, S. Yamsaki, *J. Sol. State Electrochem.*, **11**, 554 (2007).
40. Y. Hori, H. Konishi, T. Futamura, A. Murata, O. Koga, H. Sakuri, K. Oguma,

*Electrochim. Act*, **50**, 5354 (2005).

41. K. Ogura, H. Yano, T. Tanaka, *Catalysis Today*, **98**, 414 (2004).
42. a) H. Chandler, F. Pollara, *AICHE Chem. Eng. Prog. Ser.: Aerospace Life Support*, **62**, 38; b) L. Elikan, D. Archer, R. Zahradnik, *ibid*, **28** (1966).
43. a) M. Stancati, J. Niehoff, W. Wells, R, Ash, *AIAA*, *79-0906*, **262** (1979), b) R. Richter, *ibid*, *82-2275*, 1 (1981).
44. a) J. Mizusaki, H. Tagawa, Y. Miyaki, S. Yamauchi, K. Fueki, I. Koshiro, *Solid State Ionics*, *126*, 53 (1992); b) G. Tao, K. Sridhar, C. Chan, *ibid*, **175**, 615 (2004); c) *ibid*, 621; R. Green, C. Liu, S. Adler, d) *ibid*, *179*, 647 (2008).
45. C. Meyers, N. Sullivan, H. Zhu, R. Kee, *J. Electrochem. Soc.*, **158**, B117 (2011).
46. P. Kim-Lohsoontorn, N. Laosiripojana, J. Bae, *Current Appl. Physics*, **11**, 5223 (2011).
47. S. Ebbesen, C. Graves, A. Hausch, S. Jensen, M. Mogensen, *J. Electrochem. Soc.*, **157**, B1419 (2010).
48. S. Jensen, X. Sun, S. Ebbesen, R. Knibbe, M. Mogensen, *Intl. J. Hydrogen Energy*, *35*, 9544 (2010).
49. Q. Fu, C. Mabilat, M. Zahid, A. Brisse, L. Gautier, *Energy Environ. Sci.*, **3**, 1382 (2010).
50. C. M. Stoots, J. E. O'Brien, K. G. Condie, J. Hartvigsen, *Intl. J. Hydrogen Energy*, *35*, 4861 (2010).
51. Q. Fu, C. Mabilat, M. Zahid, A. Brisse, L. Gautier, *Energy Environ. Sci.*, **3**, 1382 (2010).
52. D. Lueck, W. Buttner, J. Surma, *Fluid System Technologies* (2002), at:
http://rtreport.ksc.nasa.gov/techreports/2002report/600%20Fluid%20Systems/609.html
53. S. Licht, *Electrochem. Comm.*, **4**, 789 (2002).
54. S. Licht, *J. Phys. Chem. B*, *107*, 4253 (2003).
55. S. Licht, L. Halperin, M. Kalina, M. Zidman, N. Halperin, *Chem. Comm.*, **3006** (*2003)*.
56. S. Licht, *Chem. Comm.*, **2005**, 4623 (2005).
57. A. Fujishima, K. Honda, *Nature*, **238**, 37 (1972).
58. Z. Zou, Y. Ye, K. Sayama, H. Arakawa, *Nature*, **414**, 625 (2001).
59. S. Licht, B. Wang, S. Mukerji, T. Soga, M. Umento, H. Tributsh, *J. Phys. Chem. B*, **104**, 8920 (2000),
60. S. Licht, *J. Phys. Chem. B*, **105**, 6281 (2001).
61. a) A. J. deBethune, T. S. Licht, *J. Electrochem. Soc.*, **106**, 616 (1959); b: M. W. Chase; b) *J. Phys. Chem. Ref. Data*, **9**, 1 (1998); data available at:
http://webbook.nist.gov/chemistry/form-ser.html

62. W. E. Wentworth, E. Chen, *Solar Energy*, **18**, 205 (1976).

63. J. O'M. Bockris, *Energy Options,* Halsted Press, NY **(**1980).

64. T. S. Light, S. Licht, A. C. Bevilacqua, *Electrochem & Sol State Lett.*, **8**, E16 (2005).

65. T. S. Light, S. Licht, *Anal. Chem.*, **59**, 2327 (1987).

66. S. Licht, *Anal. Chem.*, **57**, 514 (1985).

67. S. Licht, K. Longo, D. Peramunage, F. Forouzan, *J. Electroanal. Chem.*, **318**, 119 (1991).

68. C. Elschenbroich, A. Salzer *Organometallics*. 2nd Ed., Wiley-VCH, Weinheim), Germany (1992).

69. K. Sunmacher *Molten Carbonate Fuel Cells*, Wiley-VCH, Weinheim), Germany (2007).

70. J. L. Pellegrino *Energy & Environmental Profile of the U.S. Chemical Industry* (2000), available online at:

http://www1.eere.energy.gov/industry/chemicals/tools_profile.html

71. a) R.R. King, D. C. Law, K. M. Edmonson, C. M. Fetzer, G. S. Kinsey, H. Yoon, R. A. Sherif, N. H. Karam, *Appl. Phys. Lett.*, **90**, 183516 (2007); b) M. Green, K. Emery, Y. Hishikawa, W. Warata, *Prog. Photovoltaics*, **19**, 84 (2011).

72. J. E. Miller, J. E.; Allendorf, M. D.; Diver, R. B.; Evans, L. R.; Siegel, N. P.; Stuecker, J. N. *J. Mat. Sci.*, **43**, 4714 (2008).

73. Y. Woolerton, Y., W.; Sheard, S.; Reisner, E.; Pierce, E.; Ragsdale, S. W.; Armstrong, F. A. *J. Amer. Chem. Soc.*, **132**, 2132 (2010).

74. E. Benson, C. P. Kubiak, A. J, Sathrum, J. M. N. Smieja *Chem. Soc. Rev.*, **38**, 89 (2009).

75. *Solar Hydrogen Generation of: Towards a Renewable Energy Future*, Monograph, 8 chapters, Editors: K. Rajeshwar, R. McConnell, S. Licht, ***Wiley Press***, (2008).

76. S Licht, S Liu, B. Cui, J. Lau, L. Hu, J. Stuart, B. Wang, O. El-Gazawi, F.-F. Li, *J. Electrochem. Soc,* **163,** F1168 (2016); open access at:

http://jes.ecsdl.org/content/163/10/F1162.full.pdf

77. *Principles and Applications of Molten Salt Electrochemistry*, (Eds: Z. Zhang, Z. Wang) Chemical Industry Press, Beijing) p. 191 (2006).

78. T. Kojima, Y. Miyazaki, K. Nomura, K. Tanimoto, K. Density, *J. Electrochem. Soc.*, **155**, F150 (2008).

79. V. Kaplan, E. Wachtel, K. Gartsman, Y. Feldman, I. Lubormirsky, *J. Electrochem. Soc.*, **157**, B552 (2010).

80. S. Licht, B. Cui, B. Wang, *Journal of CO$_2$ Utilization,* **2**, 63 (2013).

81. J. Ren, F. Li, J. Lau, L. Gonzalez-Urbina, S. Licht, Nano Lett. **15**, 6142 (2015).
82. J. Ren, J. Lau, M. Lefler, S. Licht, Journal of Physical Chemistry, C, **119**, 23349 (2015).
83. J. Ren, S. Licht, Scientific Rep. 6, **27760** 1-11 (2016).
84. J. Ren, M. Johnson, R. Singhal, S. Licht, J. CO2 Utilization **18**, 335 (2017).
85. H. Wu, Z. Li, D. Ji, Y. Liu, L. Li, D. Yuan, Z. Zhang, J. Ren, M. Lefler, M, B. Wang, S. Licht. Carbon. **106**, 208-217 (2016).
86. S. Licht, A. Douglas, J. Ren, R. Carter, M. Lefler, CL. Pint, ACS Central Science. **2**, 162-168 (2016).
87. M. Johnson, J. Ren, M. Lefler, G. Licht, J. Vicini, J., S. Licht, Data in Brief, **14**, 592 (2018).
88. J. Lau, G. Dey, S.Licht, Energy Conservation and Management, **122**, 400 (2016).
89. S. Licht, J. CO2 Utilization, **18**, 378 (2017).
90. G. Dey, J. Ren, T. El-Ghazawi, S. Licht, RSC. Adv., **6**, 27191 (2016).
91. M. Johnson, J. Ren, M. Lefler, G. Licht, J. Vicini, X. Liu, S. Licht, Materials Today Energy, **5**, 23 (2017).
92. F.-F. Li, S. Liu, B. Cui, J. Lau, J. Stuart, S. Licht, *Advanced Energy Materials,* **7**, 1401791 (2015) with 2 page Supporting Information at:
http://onlinelibrary.wiley.com/store/10.1002/aenm.201401791/asset/supinfo/aenm201401
791- sup-0001-S1.pdf?v=1&s=987c46bbd222b740fa6923a6b69a0e2ea3607433
93. F.-F. Li, J. Lau, S. Licht, *Advanced Science*, **2**, 1500260 1-5, (2015).
94. H. Wu, D. Ji, L. Li, D. Yuan, Y. Zhu, B. Wang, Z. Zhang[*], S. Licht, Advanced Materials Technology, **1**, 60092 (2016.
95. S. Licht, B. Cui, B. Wang, F. Li, J. Lau, S. Liu, *Science*, **345**, 637-640, with 15 page online supplementary information (2014) at:
http://www.sciencemag.org/content/suppl/2014/08/06/345.6197.637.DC1.html%20
96. F.-F. Li, S. Licht, *Inorg. Chem.*, **53**, 10042 (2014), with 4 pages of supporttinh information at:
http://pubs.acs.org/doi/suppl/10.1021/ic5020048
97. B. Cui, J. Zhang, S. Liu, X. Liu, W. Xiang, L. Liu, H. Xin, M. J. Lefler S. Licht, *Green Chemistry*, **19**, 298 (2017); first published on 23 Nov 2016. pdf at:
http://pubs.rsc.org/en/content/articlepdf/2017/GC/C6GC02386J
98. S. Licht, H. Wu, C. Hettige, B. Wang, J. Lau, J. Asercion, J. Stuart, Chemical Communications, **48**, 6019, with online 20 page supplement (2012).
99. S. Licht, *J. CO2 Utilization*, **18**, 378-389 (2017). March, 19. 2017.

100. B. Wang, H. Wu, G. Zhang, S. Licht, *ChemSusChem,* **5**, 2000 (2012).

101. B. Wang, Y. Hu, H. Wu, S. Licht, *Electrochemical Science Letters,* **2**, H34 (2013).

102. Y. Zhu, B. Wang, X. Liu, H. Wang, H. Wu S. Licht, Green Chemistry, **16,** 4758, with 2 page online supplementary information (2014) at:

http://www.rsc.org/suppdata/gc/c4/c4gc01448k/c4gc01448k1.pdf

103. Y. Zhu, B. Wang, H. Wang, X. Liu, S. Licht, *Solar Energy,* **113,** 303 (2015).

104. Y. Zhu, H. Wang, B. Wang*, X. Liu, H. Wu, S. Licht*, *Applied Catalysis* B, **193**, 151 (2016).

105. B. Cui, S. Licht, *Green Chemistry*, **15** (4), 881, with 16 page online supplementary information (2013).

106. F.-F. Li, B. Wang, S. Licht*, *J. Sustainable Metallurgy*, **2**, 405 (2016). Cover article, available open

access http://link.springer.com/article/10.1007/s40831-016-0062-8

107. L. Andrieux, G. Weiss, *Comptes Rendu*, **217**, 615 (1944).

108. G. M. Haarberg, E. Kvalheim, S. Rolseth, T. Murakami, S. Pietrzyk, S. Wang, *ECS Transactions*, **3**, 341 (2007).

109. S. Wang, G. M. Haarberg, E. Kvalheim, E. *J. Iron and Res. Int.*, **15**, 48 (2008).

110. G. M. Li, D. H. Wang, Z. Chen, *J. Mat. Sci. Tech.*, **25**, 767 (2009).

111. B. Y. Yuan, O. E. Kongstein, G. M. Haarberg, *J. Electrochem. Soc.*, **156**, D64 (2009).

112. W. Palmaer, J. A. Brinell, *Chem. Metall. Eng.*, **11**, 197 (1913).

113. F. A. Eustis, *Chem. Metall. Eng.*, **27**, 684 (1922).

114. E. Mostad, S. Rolseth, S. Thonstad, J. *Hydrometallurgy*, **90**, 213 (2008).

115. L. Qingeng, F. Borum, I. Petrushina, N. J. Bjerrum, *J. Electrochem, Soc.*, **146**, 2449 (1999.

116. R. Collongues, G. Chaudron, *Compt. Rend.*, **124**, 143 (1950).

117. A. Wijayasinghe, B. Bergman, C. Lagergren, *J. Electrochem. Soc.*, **150**, A558 (2003).

118. A. Lykasov, M. Pavlovskaya, *Inorg. Mat.*, **39**, 1088 (2003).

119. H. Q. Li, S. S. Xie, *J. Rare Earths*, **23**, 606 (2005).

120. G. Demirci, I. Karakaya, *J. Alloys & Compounds*, **465**, 255 (2008).

121. C. S. Solanki, G. Beaucarne, *Advanced Solar Cell Concepts*, AER India-2006, **256** (2006).

122. A. Luque, A. Marti, *Handbook of Photovoltaic Sci. & E*ng., (Eds. A. Luque, S.

Haegedus), Wiley-VCH, Weinheim, Germany, 113 (2003).

123. A. Kogan, *Intl. J. Hydrogen Energy*, **23**, 89 (1998).

124. A. Andrews, J. Logan, Fischer-Tropsch Fuels from Coal, Natural Gas, and Biomass: Background and Policy. *Congressional Research Service Report for Congress*, RL34133, (2008), (March 27, 2008); available at: http://assets.opencrs.com/rpts/RL34133_20080327.pdf.

125. E. Barbier, How is the global green deal going? *Nature*, **464**, 832 (2010).

126. Power tower solar technologies are described at:

brightsourceenergy.com; ausra.com, esolar.com; bengoasolar.com/corp/web/en/our_projects/solana/

127. Siemens to build molten salt solar thermal test facility in Portugal, siemens.com, (2011), at:

http://www.siemens.com/press/pool/de/pressemitteilungen/2011/renewable_energy/ERE201102037e.pdf

128. solarreserve.com, at: http://www.solarreserve.com/projects.html. (2011).

129. Parabolic solar concentrator technologies are described at: stirlingenergy.com.

130. Fresnel solar concentrator technologies are described at: amonix.com, energyinnovations.com/sunflower.

131. R. Pitz-Paal, Solar Energy Conversion and Photoenergy Systems, Eds. Galvez, J. B.; Rodriguez, S. M., *EOLSS* Publishers, Oxford, UK. (2007).

132. P. Tans, *Oceanography*, **22**, 26 (2009).

133. Tahil, W., 54 pages, Meridan International Research, Martainsville, France (2008).

**Figures and Schemes with captions**

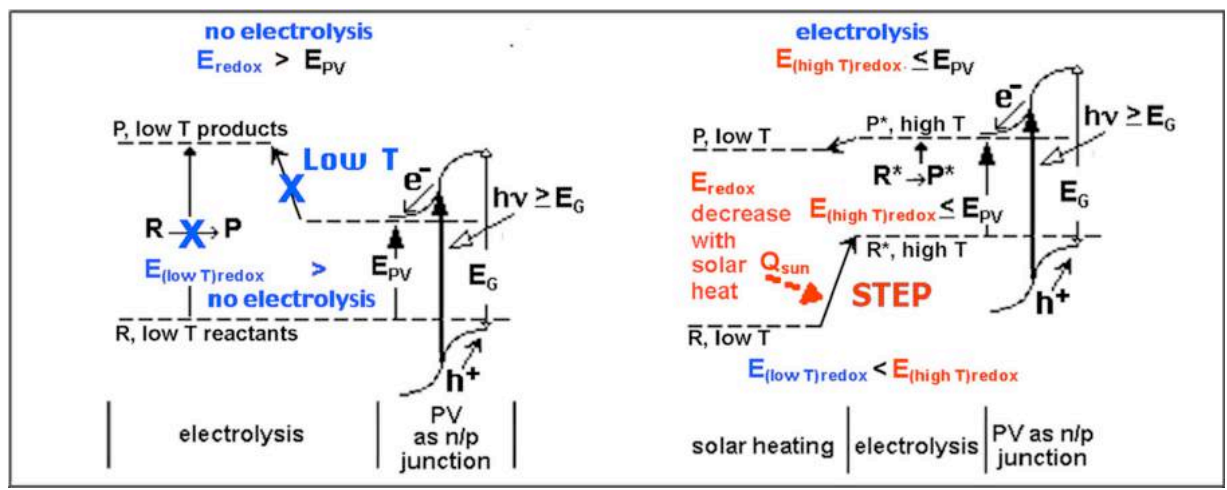

**Scheme 1.** Top: Comparison of PV and **STEP** solar driven electrolysis energy diagrams. STEP uses sunlight to drive otherwise energetically forbidden pathways of charge transfer. The energy of photodriven charge transfer is insufficient (left) to drive (unheated) electrolysis, but is sufficient (right) to drive endergonic electrolysis in the solar heated synergestic process. The process uses both visible & thermal solar energy for higher efficiency; thermal energy decreases the electrolysis potential forming an energetically allowed pathway to drive electrochemical charge transfer.

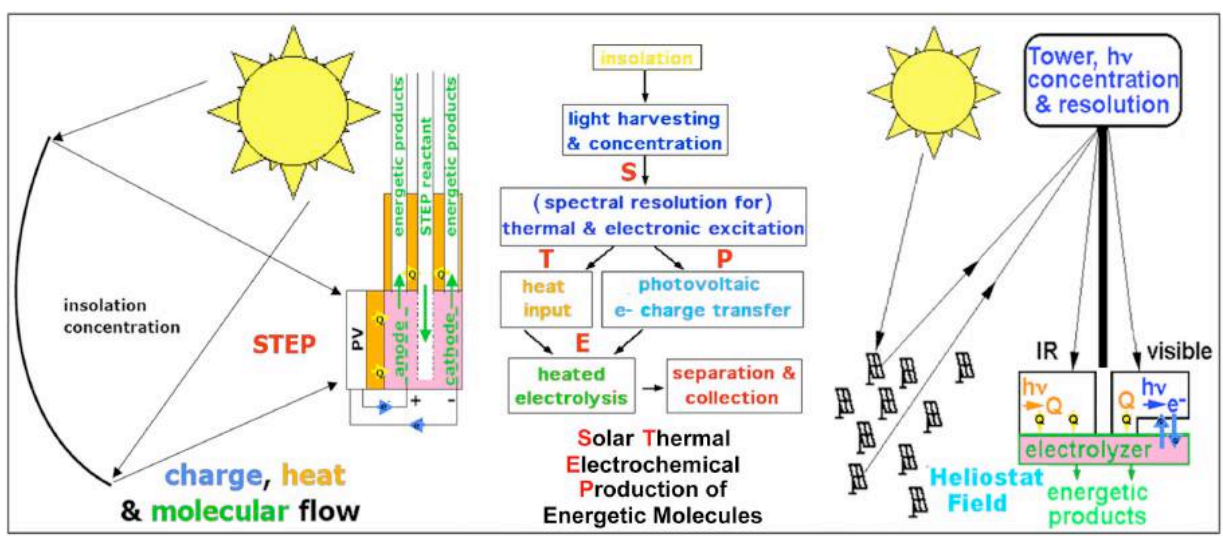

**Scheme 2.** Global use of sunlight to drive the formation of energy rich molecules. **Left**: Charge, & heat flow in **STEP**: heat flow (yellow arrows), electron flow (blue), & reagent flow (green). **Right**: Beam splitters redirect sub-bandgap sunlight away from the **PV** onto the electrolyzer.

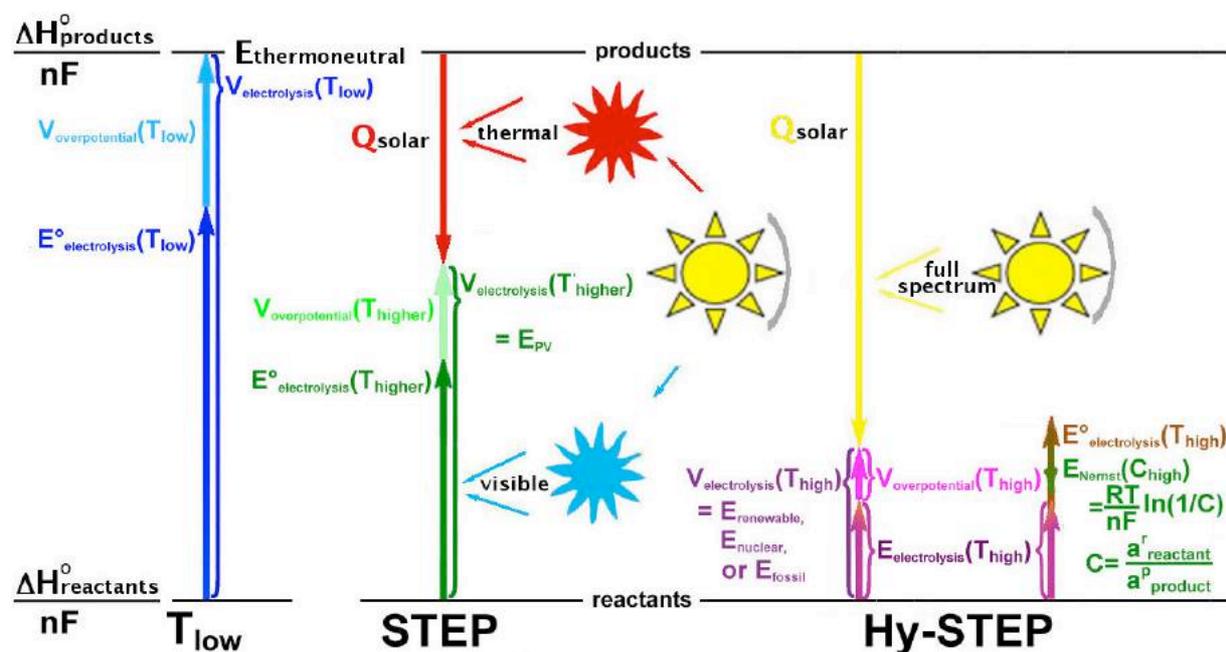

**Scheme 3**. Comparison of solar energy utilization in STEP and Hy-STEP implementations of the solar thermal electrochemical production of energetic molecules.

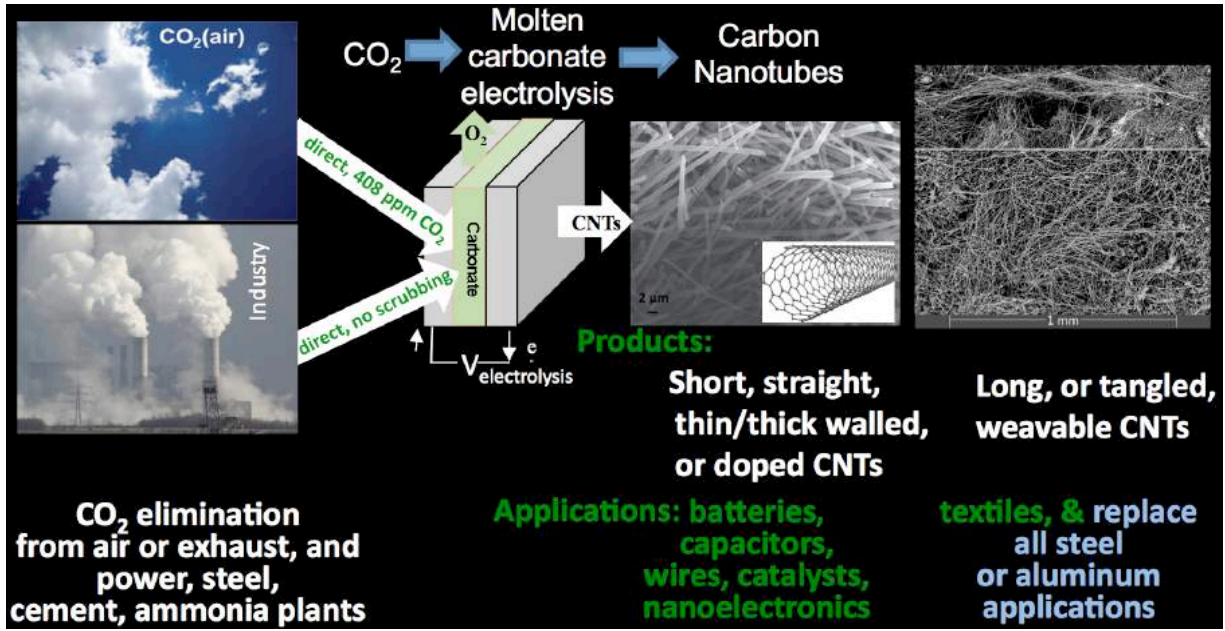

**Scheme 4.** Stronger by weight and more stress resistant than steel, C2CNT carbon nanotubes, are fomed from carbon to mitigate this greenhouse gas by molten carbonate electrolysis. Controlled electrolysis parameters have led to a wide portfolio of uniform short or long, thin or thick, straight or tangled with defects, and doped or undoped C2CNT CNTs.[81-91]

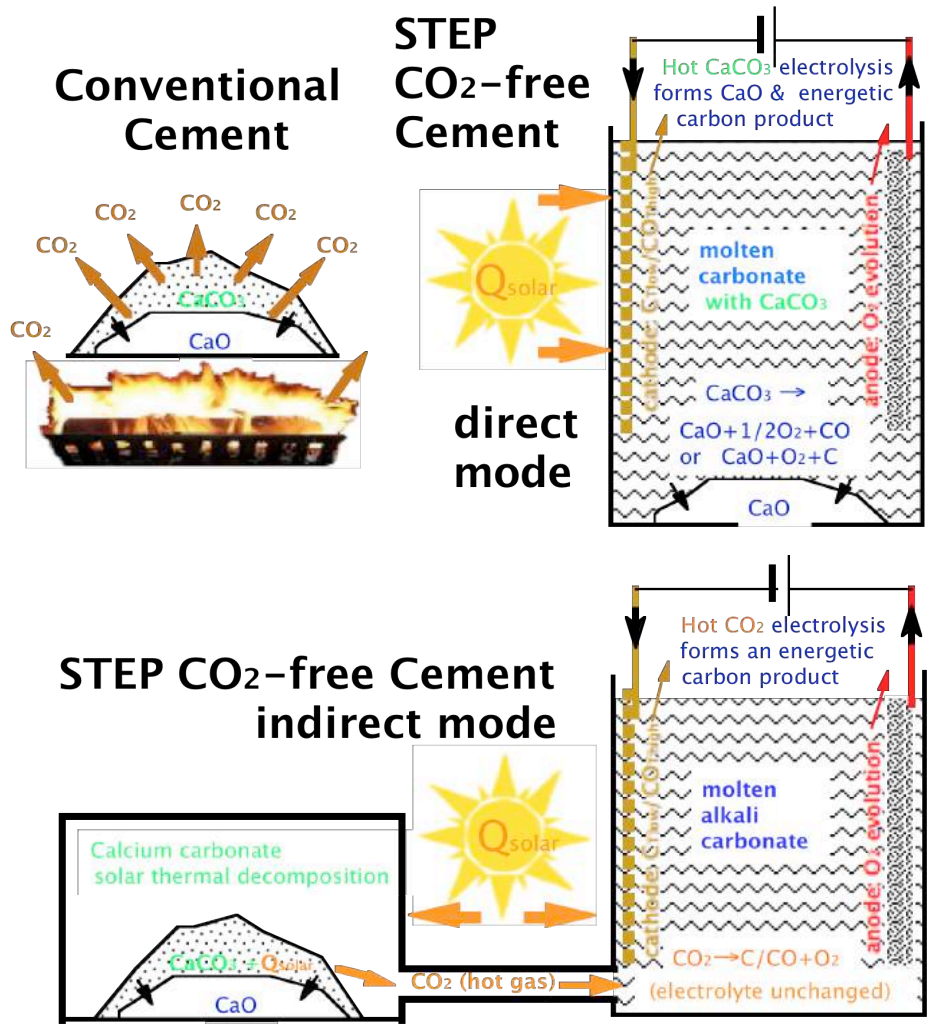

**Scheme 5. Top left**: Conventional production of lime for cement by the thermal decomposition of CaCO3. **Top right:** STEP direct solar conversion of calcium carbonate to lime (top right) eliminating $CO_2$. **Bottom:** the indirect mode of STEP cement.[98]

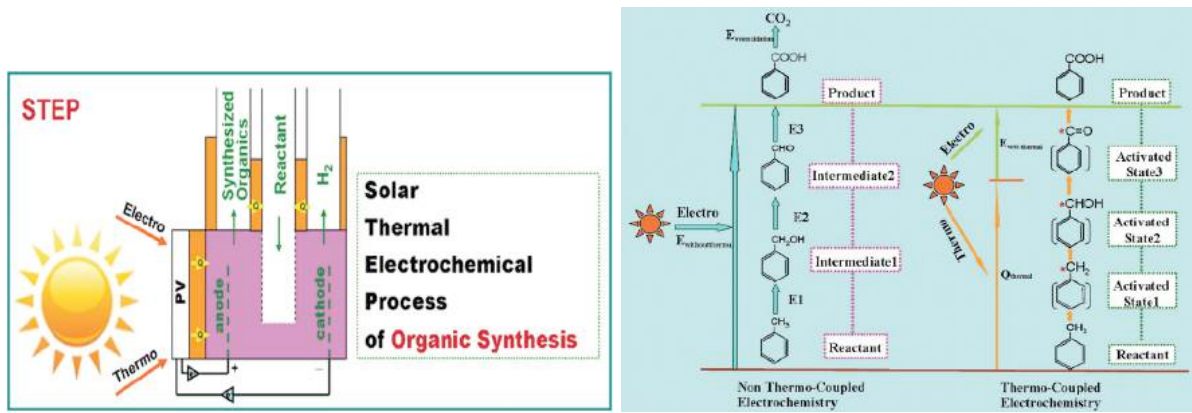

**Scheme 6. Left**. STEP organic. **Right**. Solar thermal activiation of the electrolysis of toluene to benzoic acid.[104]

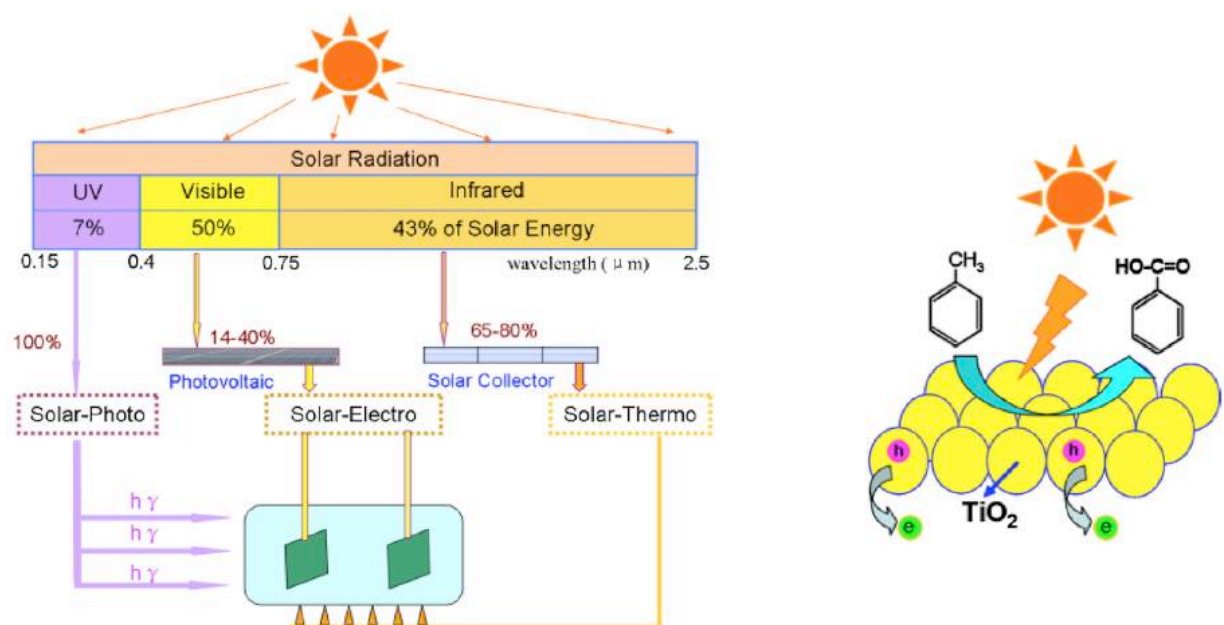

**Scheme 7 Left**. Simultaneous application and synergism of solar thermal, solar electro and solar photocatalytic activation and enhanmenent of STEP organic. **Right**. Photocatalystic oxidation of toluene to benzoic acid at the $TiO_2$ electrode; preparation of the high activity $TiO_2$ photoelectrodes is detailed in reference.[104]

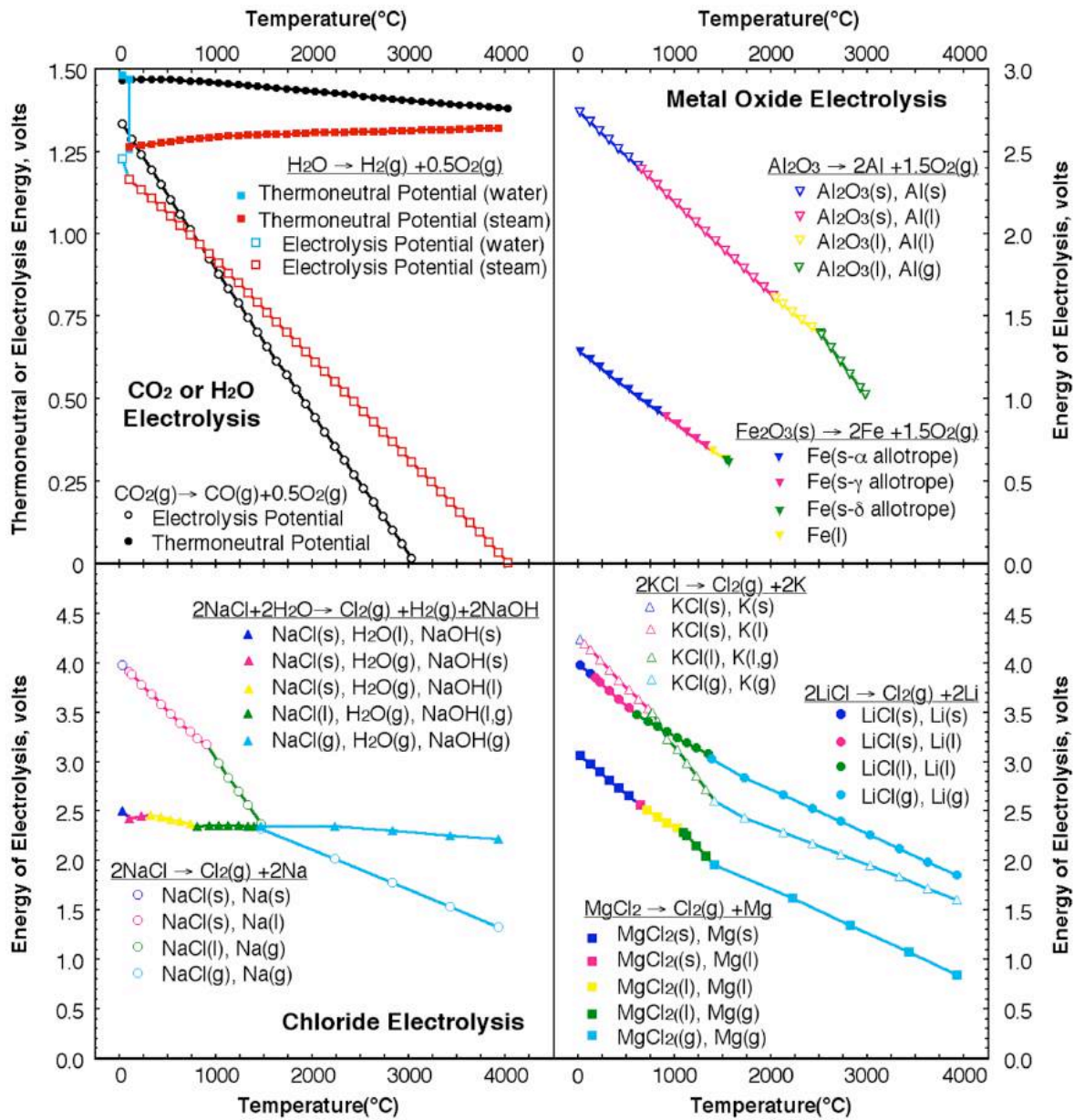

**Figure 1.** The calculated potential to electrolyze selected oxides (top) and chlorides (bottom). The indicated decrease in electrolysis energy, with increase in temperature, provides energy savings in the **STEP** process in which high temperature is provided by excess solar heat. Energies of electrolysis are calculated from eq 3, with consistent thermochemical values at unit activity using NIST gas and condensed phase Shomate equations.[59b] Note with water excluded, the chloride electrolysis decreases (in the lower left of the figure). All other indicated electrolysis potentials, including that of water or carbon dioxide, decrease with increasing temperature. Thermoneutral potentials are calculated with eq 5.

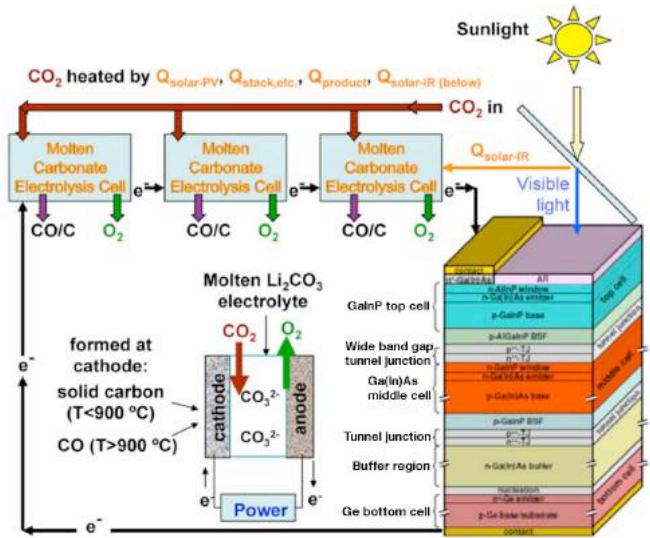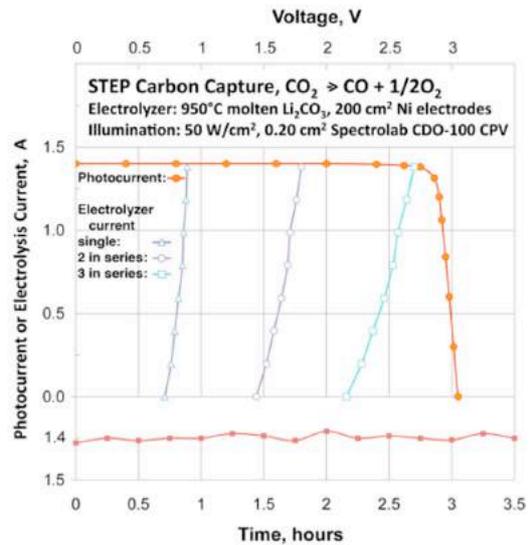

**Figure 2**. Left: STEP carbon capture in which three molten carbonate electrolysis in series are driven by a concentrator photovoltaic. Sunlight is split into two spectral regions; visible drives the CPV and thermal heats the electrolysis cell. In Hy-STEP (not shown) sunlight is not split and the full spectrum heats the electrolysis cell, and electronic charge is generated separately by solar, wind, or other source. Right: The maximum power point photovoltage of one Spectrolab CPV is sufficient to drive three in series carbon dioxide splitting 950°C molten $Li_2CO_3$ electrolysis cells. Top: Photocurrent at 500 suns (masked (0.20 cm$^2$) Spectrolab CDO-100 CPV, or electrolysis current, versus voltage; electrolysis current is shown of one, two or three series 950°C $Li_2CO_3$ electrolysis cells with 200 cm$^2$ Ni electrodes. Three in series electrolysis cells provide a power match at the 2.7 V maximum power point of the CPV at 950°C; similarly (not shown), two 750°C $Li_2CO_3$ electrolysis cells in series provide a power match at 2.7V to the CPV. Bottom: Stable carbon capture (with 200 cm$^2$ "aged" Ni electrodes at 750°C; fresh electrodes (not shown) exhibit an initial fluctuation as carbon forms at the cathode and Ni oxide layer forms on the anode. The rate of solid carbon deposition gradually increases as the cathode surface area slowly increases in time.

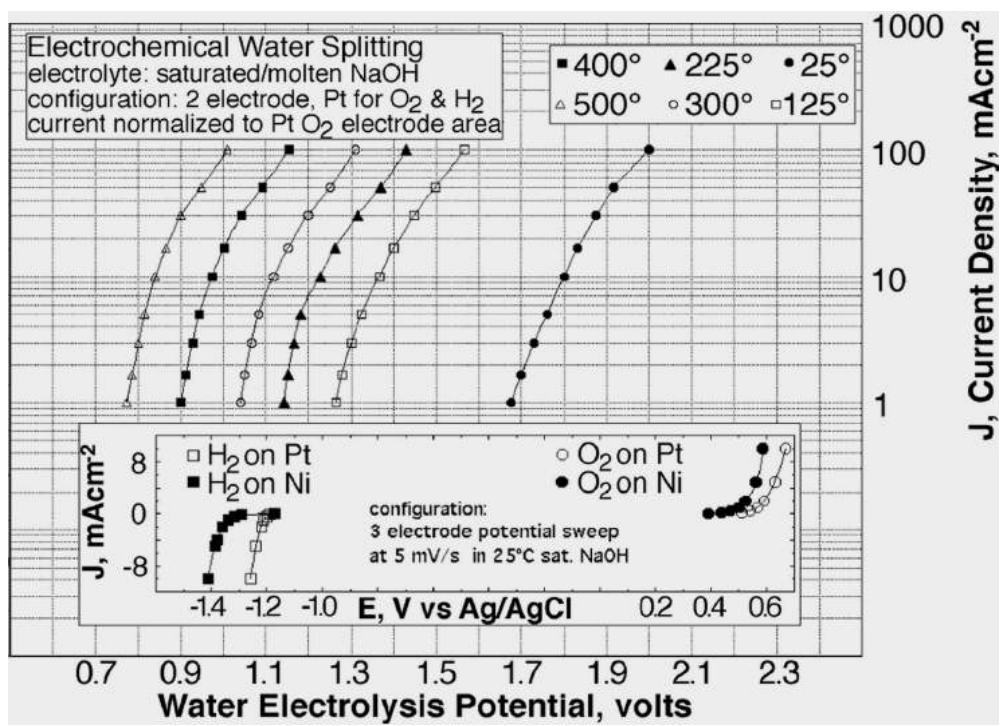

**Figure 3**. The water electrolysis potential measured in aq. saturated or molten NaOH, at 1 atm. Steam is injected in the molten electrolyte. $O_2$ anode is 0.6 cm$^2$ Pt foil. IR and polarization losses are minimized by sandwiching 5 mm from each side of the anode, oversized Pt gauze cathode. Inset: At 25C°, 3 electrode values comparing Ni and Pt working electrodes and with a Pt gauze counterelectrode at 5 mV/s.

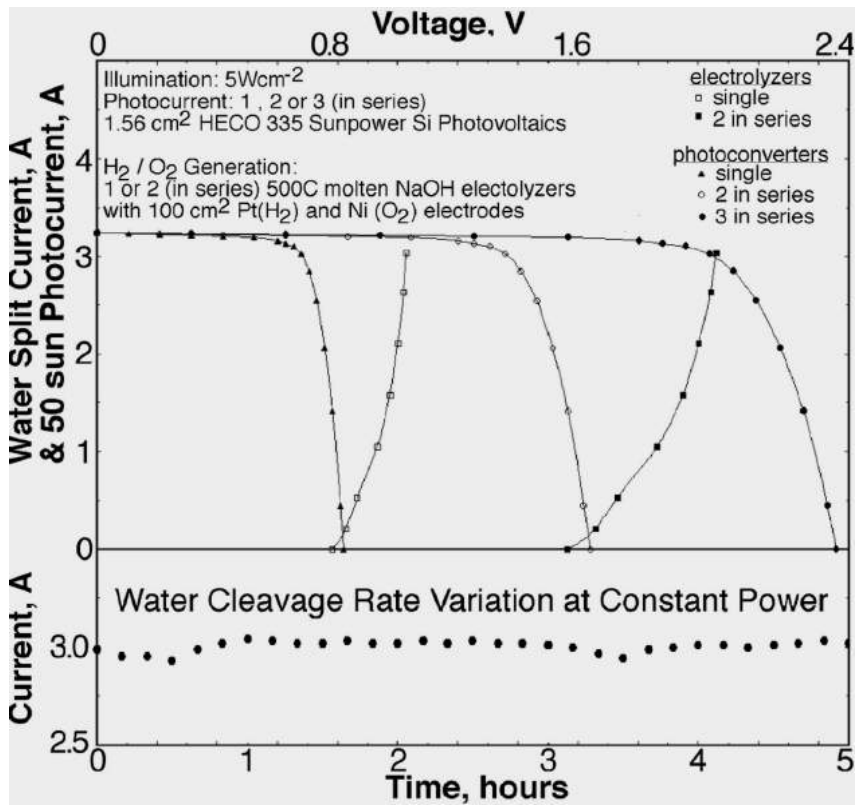

**Figure 4**. Photovoltaic and electrolysis charge transfer of STEP hydrogen using Si CPV's driving molten NaOH water electrolysis. Photocurrent is shown for 1, 2 or 3 1.561 cm$^2$ HECO 335 Sunpower Si photovoltaics in series at 50 suns. The CPV's drive 500°C molten NaOH steam electrolysis using Pt gauze electrodes. Left inset: electrolysis current stability.

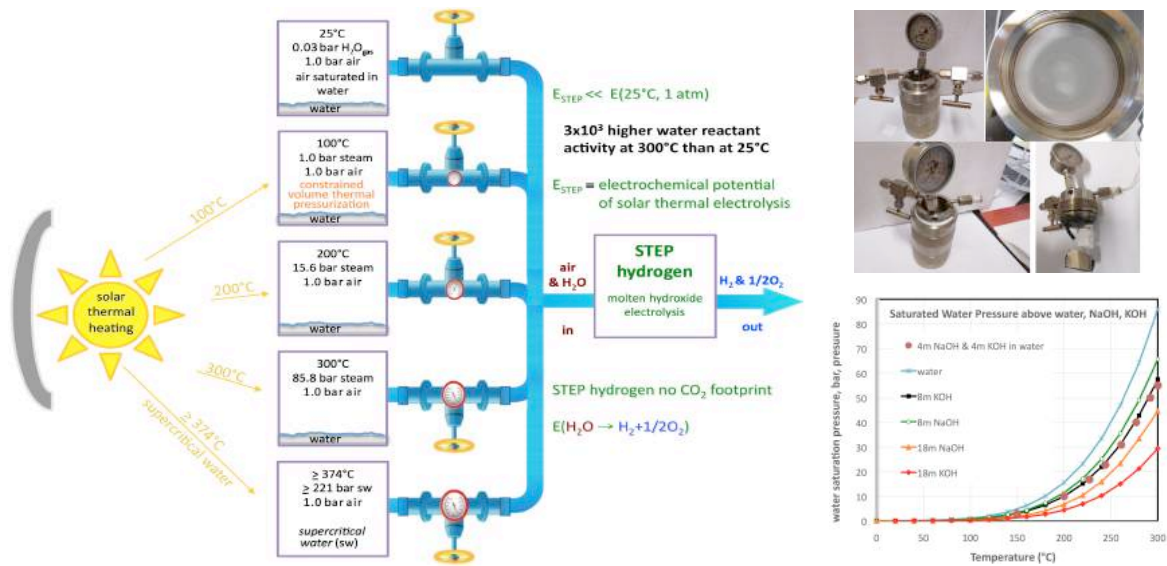

**Figure 5**. **Right:** The pressure of water in a confined environment is enhanced with solar heating, as a feedstock for electrolytic hydrogen production. **Top right:** High-pressure electrolytic water splitting hydrogen cell in-house modified from a hydrothermal cell by replacement of a flow valve with throughput electrical contacts. **Bottom right:** measured water pressure above a 4 molal NaOH + 4 molal KOH aqueous solution (a mix containing a 1:1 molar ratio of NaOH to KOH and 72.2 mass percent water) compared to the known experimental saturated water pressures above water, 8 or 18 m NaOH, and 8 or 18 m KOH.[76]

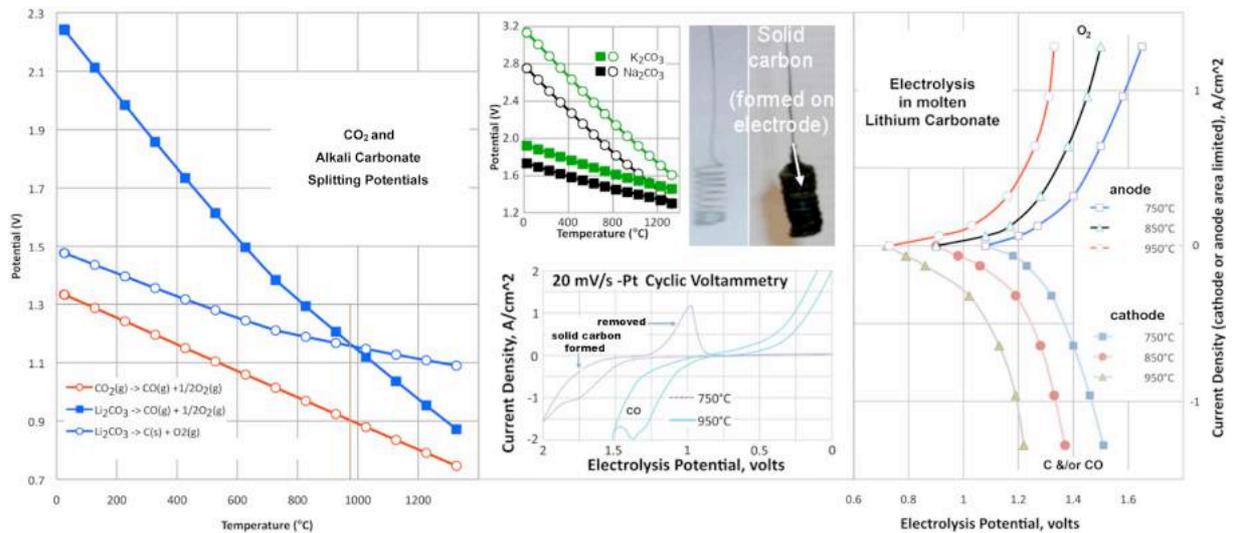

**Figure 6**. The calculated (left) and measured (right) electrolysis of $CO_2$ in molten carbonate. Left: The calculated thermodynamic electrolysis potential for carbon capture and conversion in $Li_2CO_3$ (main figure), or $Na_2CO_3$ or $K_2CO_3$ (left middle); squares refer to $M_2CO_3 \rightarrow C + M_2O + O_2$ and circles to a $M_2CO_3 \rightarrow CO + M_2O + 1/2O_2$. To the left of the vertical brown line, solid carbon is the thermodynamically preferred (lower energy) product. To the right of the vertical line, CO is preferred. Carbon dioxide fed into the electrolysis chamber is converted to solid carbon in a single step. Photographs: coiled platinum cathode before (left), and after (right), $CO_2$ splitting to solid carbon at 750 °C in molten carbonate with a Ni anode. Right: The electrolysis full cell potential is measured, under anode or cathode limiting conditions, at a platinum electrode for a range of stable anodic and cathodic current densitites in molten $Li_2CO_3$. Lower midde: cathode size restricted full cell cyclic voltammetry, CV, of Pt electrodes in molten $Li_2CO_3$.

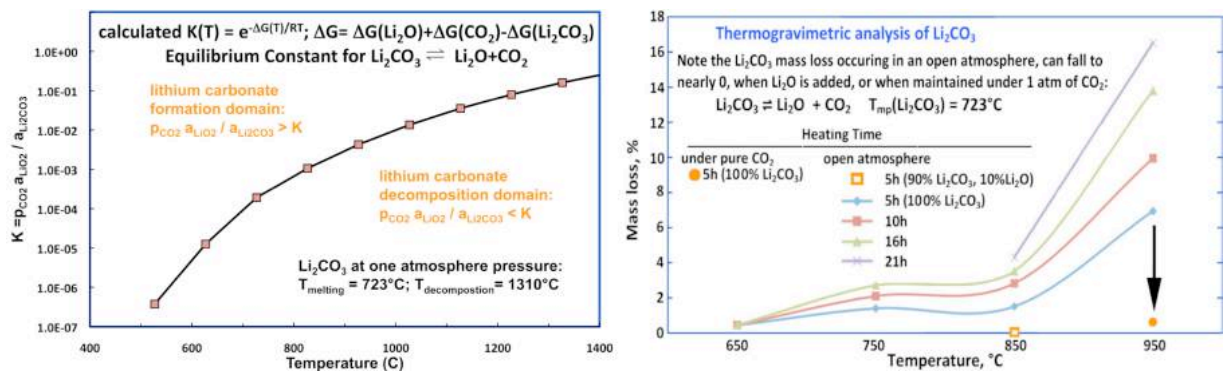

**Figure 7.** Left: Species stability in the lithium carbonate, lithium oxide, carbon dioxide system, as calculated from $Li_2CO_3$, $Li_2O$, and $CO_2$ thermochemical data. Right: Thermogravimetric analysis of lithium carbonate. The measured mass loss in time of $Li_2CO_3$.

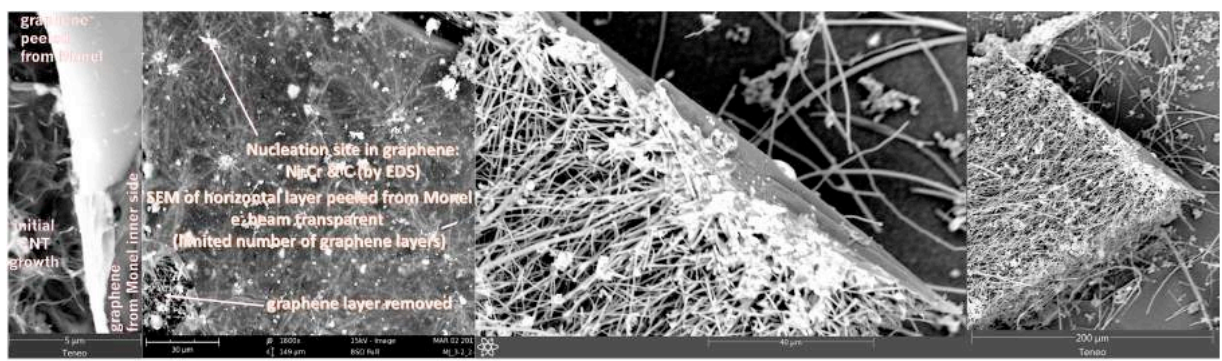

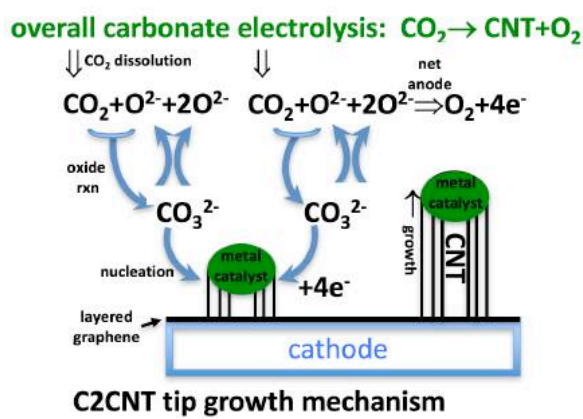

**Figure 8.** Top: SEM of product formed in the initial electrolysis stages at the Monel cathode/equilibrated electrolyte interface. Bottom: Carbonate electrolytic growth model of carbon nanotubes from $CO_2$. The mechanism is based on the layered graphene observation observed in the top of the figure, and our previous SEM, EDX, TEM, chemical balance (the oxide buildup when $CO_2$ is exclude), DFT calculations, and isotopic evidence. This proposed tube tip growth mechanism occurs at the solid/liquid (molten carbonate) interface, and transforms $CO_2$ into CNTs. The mechanism is analogous to the CVD growth mechanism, which instead occurs at the solid/gas, interface, and transforms organics, rather than $CO_2$, into CNTs.[88,91]

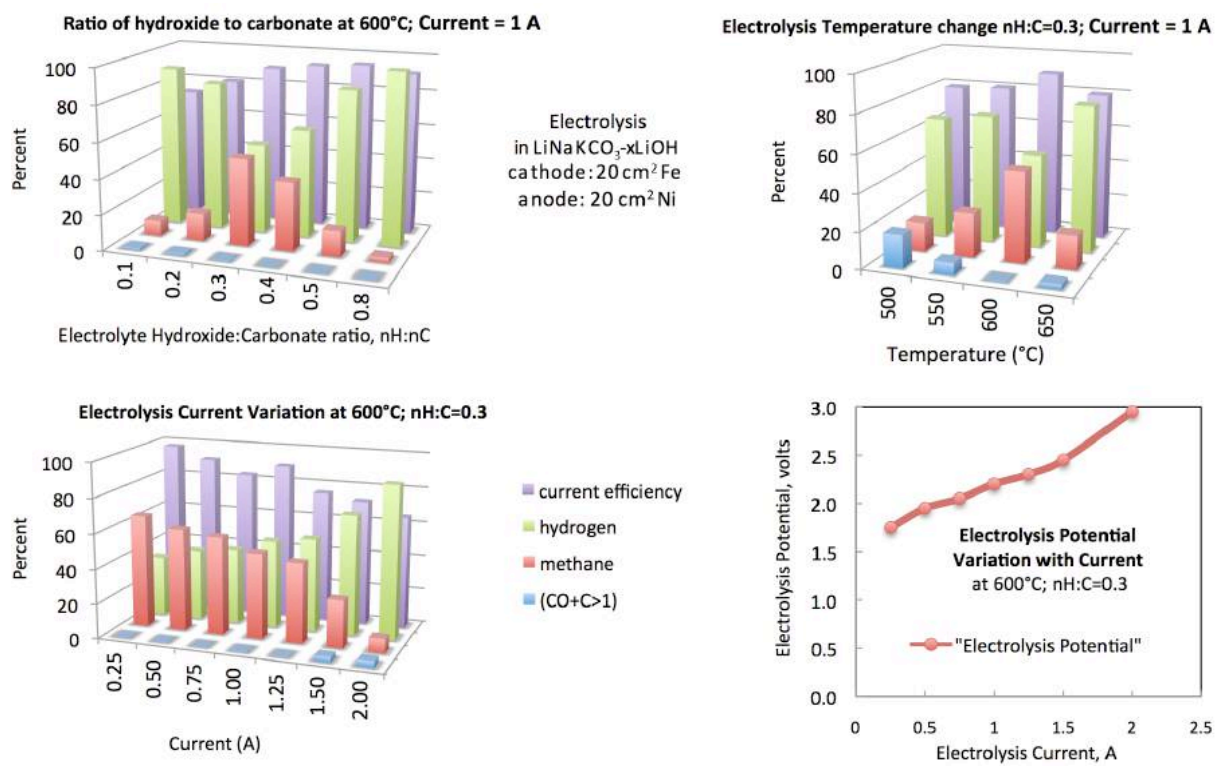

**Figure 9.** Composition of electrolysis gas products, current efficiency, and electrolysis potential in molten composite carbonate:LiOH electrolytes with an Fe cathode and an Ni anode (75c).

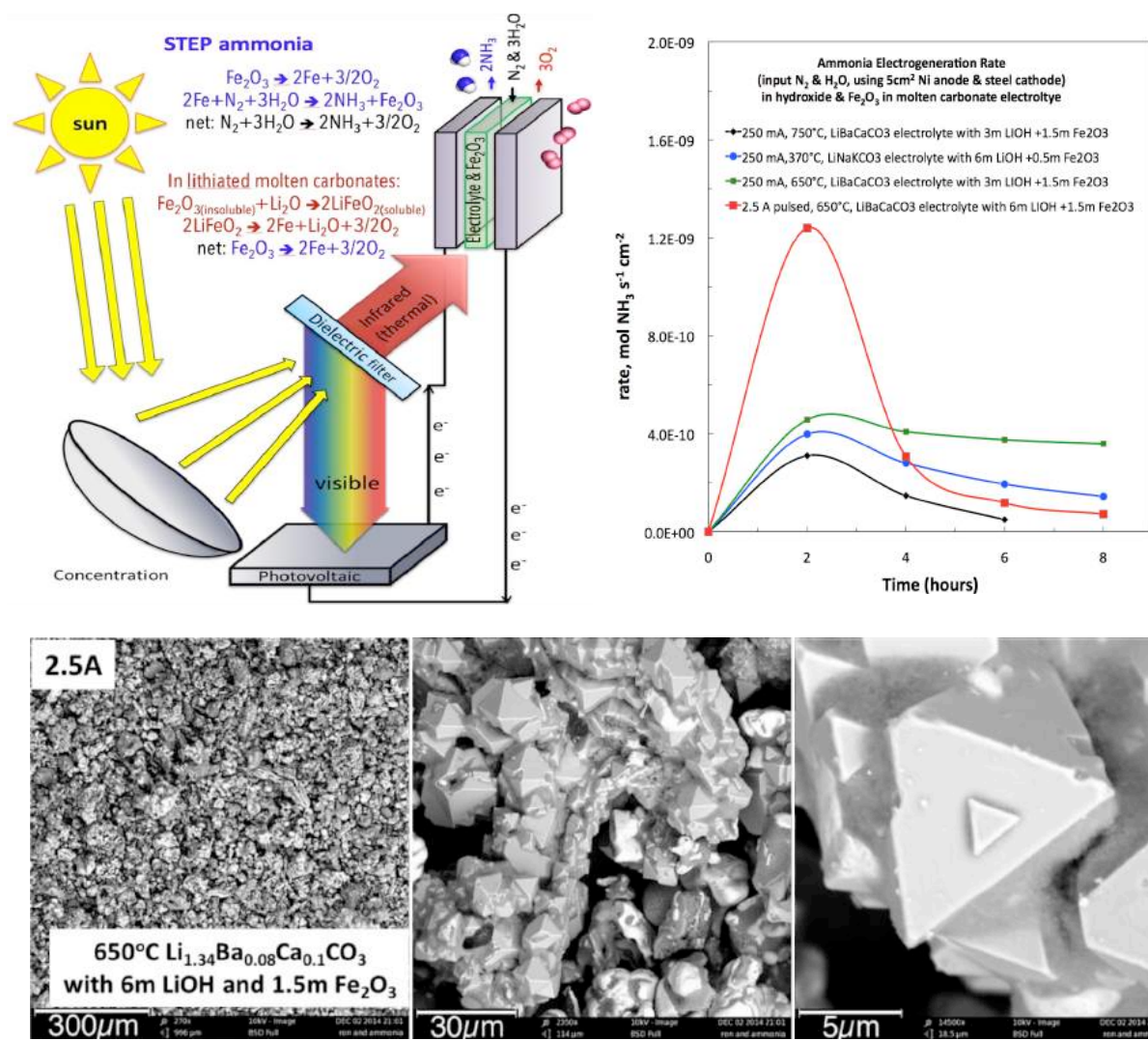

**Figure 10. Top left**. STEP ammonia. Incident sunlight is concentrated and split. The visible is incident on a solar cell, such as a CPV, which drives an electrolysis chamber heated by the solar thermal. The electrolysis drives iron formation from $Fe^{3+}$, reacting with nitrogen and water to ammonia. **Top right**. The rate of ammonia formation by electrolysis at low and intermediate temperatures in mixed hydroxide/carbonate electrolytes containing $Fe_2O_3$. **Bottom**. SEM of the iron product after 2.5A 650°C $Li_{1.34}Ba_{0.08}Ca_{0.1}CO_3$ electrolysis with 6m LiOH and 1.5m $Fe_2O_3$. iron is analyzed titrametrically and by PHENOM EDS SEM. Cathodic iron and octahedral magnetite, $Fe_3O_4$ are observed. The effect of water is evident as oxidized iron.

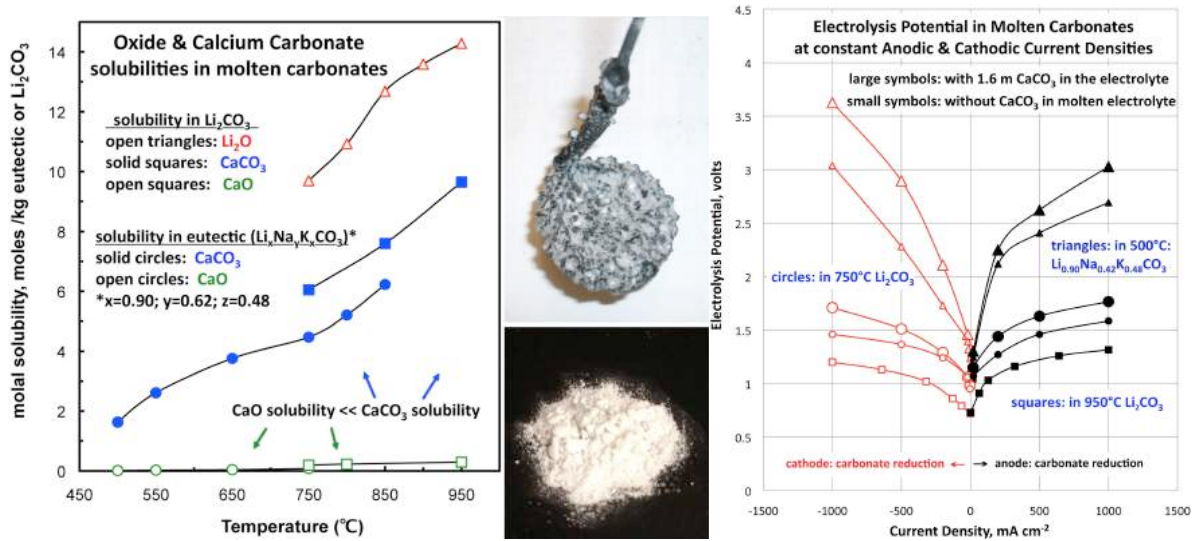

**Figure 11. Left**. The low solubility of calcium oxide, compared to calcium carbonate and lithium oxide solubility in molten carbonates facilitates the electrolysis and precipitation of calcium oxide. **Middle**. Products of the direct mode of STEP cement; carbon on the cathode (coated with congealed electrolyte after extraction) and precipated lime. **Right**. The measured full electrolysis potential as a function of current density in either $Li_2CO_3$ at 750 or 950 °C, or eutectic molten carbonates at 500 °C (76h).

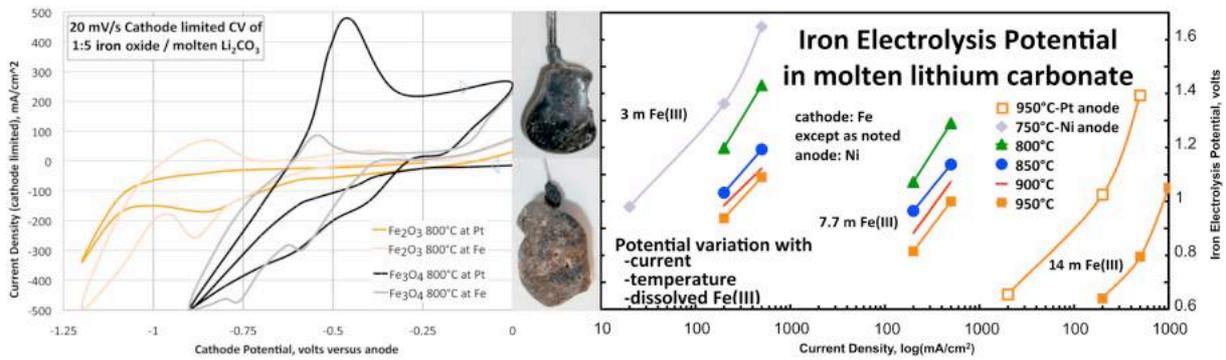

**Figure 12**. Middle: Photographs of electrolysis products from 20% $Fe_2O_3$ or $Fe_3O_4$ by mass in 800°C $Li_2CO_3$: following extended 0.5A electrolysis at a coiled wire (Pt or Fe) cathode with a Ni anode. Left: cathode restricted CV in $Li_2CO_3$, containing 1:5 by weight of either $Fe_2O_3$ or $Fe_3O_4$. Right: The measured iron electrolysis potentials in molten $Li_2CO_3$, as a function of the temperature, current density, and the concentration of dissolved Fe(III).

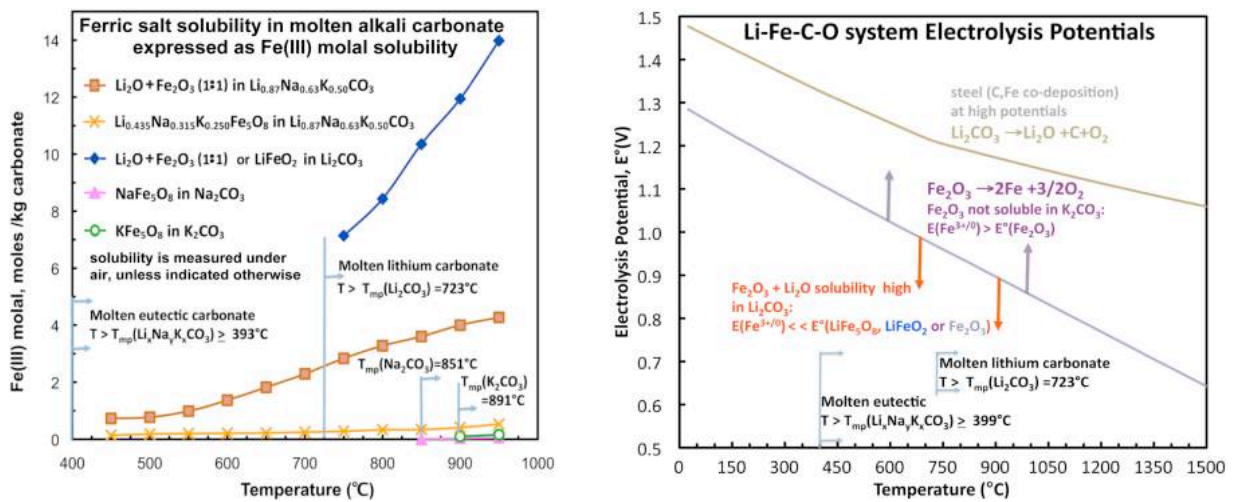

**Figure. 13.** Left: Measured ferric oxides solubilities in alkali molten carbonates. Right: Calculated unit activity electrolysis potentials of $LiFe_5O_8$, $Fe_2O_3$ or $Li_2CO_3$. Vertical arrows indicate Nernstian shifts at high or low Fe(III).

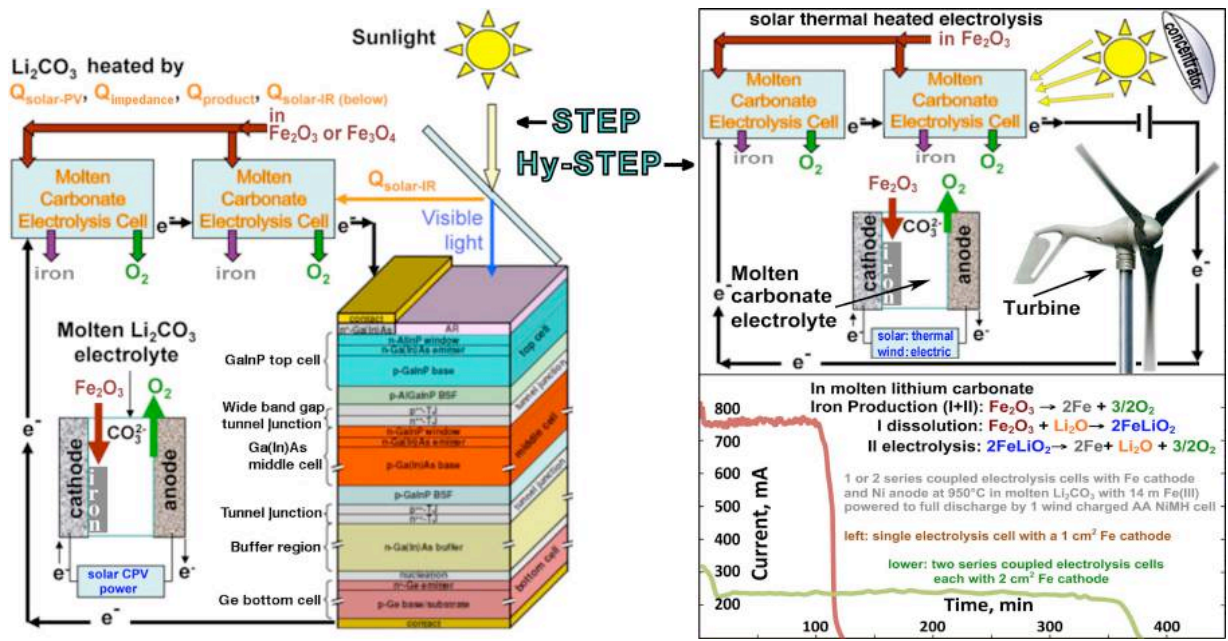

**Figure 14**. STEP and (wind) Hy-STEP iron. Left: STEP iron production in which two molten carbonate electrolysis in series are driven by a concentrator photovoltaic. The 2.7 V maximum power of the CPV can drive either two 1.35 V iron electrolyses at 800°C (schematically represented), or three 0.9 V iron electrolyses at 950°C. At 0.9V, rather than at E°(25°C) =1.28V, there is a considerably energy savings, achieved through the application of external heat, including solar thermal, to the system. Right: The Hy-STEP solar thermal/wind production of $CO_2$ free iron. Concentrated sunlight heats, and wind energy drives electronic transfer into the electrolysis chamber. The required wind powered electrolysis energy is diminished by the high temperature and the high solubility of iron oxide. Bottom: Iron is produced at high current density and low energy at an iron cathode and with a Ni anode in 14 m $Fe_2O_3$ + 14 m $Li_2O$ dissolved in molten $Li_2CO_3$.

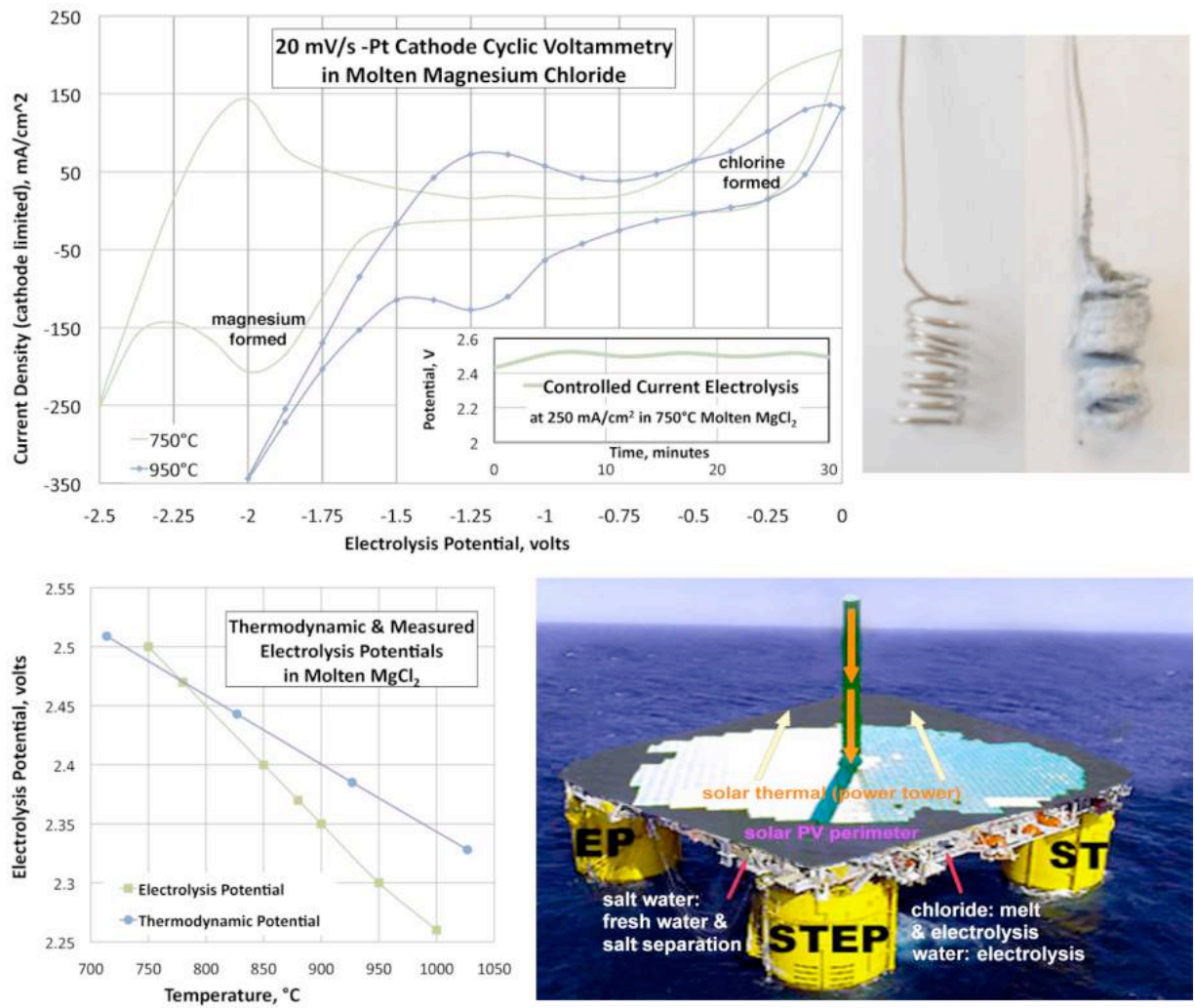

**Figure 15**. Photograph lower left: coiled platinum before (left), and after (right), $MgCl_2$ electrolysis forming Mg metal on the cathode (shown) and evolving chlorine gas on the anode. Main figure: cathode size restriced cyclic voltammetry of Pt electrodes in molten $MgCl_2$ Inset: The measured full cell potential during constant current electrolysis at 750°C in molten $MgCl_2$. Lower right: Thermodynamic and measured electrolysis potentials in molten $MgCl_2$ as a function of temperature. Electrolysis potentials are calculated from the thermodynamic free energies components of the reactants and products as E= -ΔG(reaction)/2F. Measured electrolysis potentials are stable values on Pt at 0.250 A/cm$^2$ cathode.[8] Lower right: A schematic representation of a separate (i) solar thermal and (ii) photovoltaic field to drive both water purification, hydrogen generation, and the endergonic electrolysis of the separated salts to useful products.

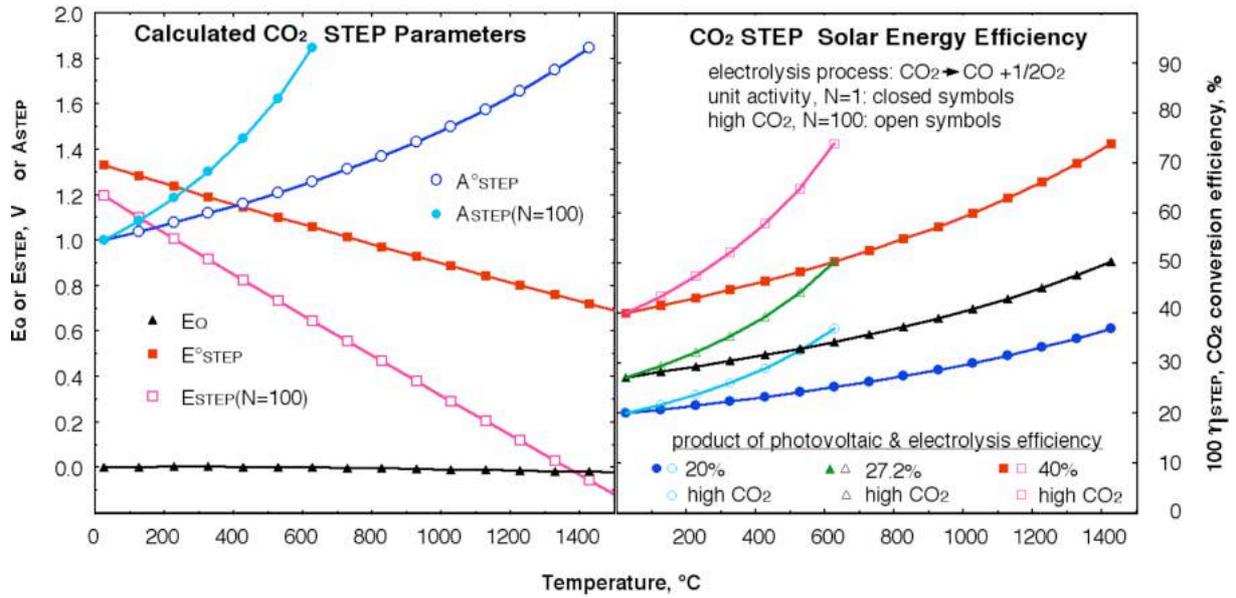

**Figure 16.** Top: Calculated **STEP** parameters for the solar conversion of $CO_2$. Bottom: Solar to chemical conversion efficiencies calculated through eq 32 for the conversion of $CO_2$ to CO and $O_2$. In the case in which the product of the photovoltaic and electrolysis efficiency is 27.2% ($\eta_{PV} \cdot \eta_{electrolysis} = 0.272$), the **STEP** conversion efficiency at unit activity is 35%, at the 650°C temperature consistent with molten carbonate electrolysis, rising to 40% at the temperature consistent with solid oxide electrolysis (1000°C). Non-unit activity calculations presented are for the case of $\sqrt{2}\, a_{CO_2}\, a_{CO}^{-3/2} = 100$. A solar conversion efficiency of 50% is seen at 650°C when N=100 (the case of a cell with 1 bar of $CO_2$ and ~58 mbar CO).

**Brief Summary:**

STEP (Solar Thermal Electrochemical Production) is derived and experimentally verified for the electrosynthesis of energetic molecules at solar energy efficiency greater than any photovoltaic conversion efficiency. In STEP the efficient formation of metals, fuels, ammonia, chlorine, and carbon capture is driven by solar thermal heated endergonic electrolyses of concentrated reactants at high solar energy conversion efficiency.

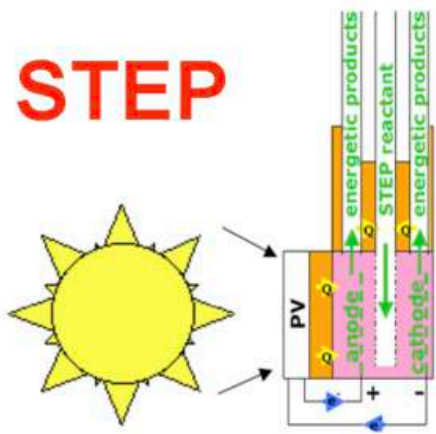